\documentclass[pdflatex,sn-mathphys-num]{sn-jnl}

\usepackage{graphicx}%
\usepackage{multirow}%
\usepackage{amsmath,amssymb,amsfonts}%
\usepackage{amsthm}%
\usepackage{mathrsfs}%
\usepackage[title]{appendix}%
\usepackage{textcomp}%
\usepackage{manyfoot}%
\usepackage{booktabs}%
\usepackage{algorithm}%
\usepackage{algorithmicx}%
\usepackage{algpseudocode}%
\usepackage{listings}%
\usepackage{todonotes}

\usepackage{multirow}
\usepackage{graphicx} 
\usepackage{diagbox}
\usepackage{makecell}
\usepackage{float}
\usepackage{colortbl}
\usepackage{xcolor}
\usepackage{array}
\usepackage{enumitem}
\usepackage{hyperref}
\usepackage{amsfonts}

\theoremstyle{thmstyleone}%
\theoremstyle{thmstyletwo}%

\theoremstyle{thmstylethree}%

\begin{document}
\title[Article Title]{Fluorescence Diffraction Tomography using Explicit Neural Fields}


\author[1]{\fnm{Renzhi} \sur{He}}\email{cubhe@ucdavis.edu}
\author[1]{\fnm{Yucheng} \sur{Li}}\email{ycsli@ucdavis.edu}
\author[1]{\fnm{Junjie} \sur{Chen}}\email{jujchen@ucdavis.edu}
\author*[1]{\fnm{Yi} \sur{Xue}}\email{yxxue@ucdavis.edu}

\affil[1]{\orgdiv{Department of Biomedical Engineering}, \orgname{University of California, Davis}, \orgaddress{\street{451 Health Sciences Dr.}, \city{Davis}, \postcode{95616}, \state{CA}, \country{United States}}}



\abstract{
Simultaneous imaging of fluorescence-labeled and label-free phase objects in the same sample provides distinct and complementary information. Most multimodal fluorescence-phase imaging operates in transmission mode, capturing fluorescence images and phase images separately or sequentially, which limits their practical application in vivo. Here, we develop fluorescence diffraction tomography (FDT) with explicit neural fields to reconstruct the 3D refractive index (RI) of phase objects from diffracted fluorescence images captured in reflection mode. The successful reconstruction of 3D RI using FDT relies on four key components: a coarse-to-fine structure, self-calibration, a differential multi-slice rendering model, and partially coherent masks. The explicit representation integrates with the coarse-to-fine structure for high-speed, high-resolution reconstruction, while the differential multi-slice rendering model enables self-calibration of fluorescence illumination, ensuring accurate forward image prediction and RI reconstruction. Partially coherent masks efficiently resolve discrepancies between the coherent light model and partially coherent light data. FDT successfully reconstructs the RI of 3D cultured label-free bovine myotubes in a 530 $\times$ 530 $\times$ 300 $\mu m^3$ volume at 1024 $\times$ 1024 pixels across 24 $z$-layers from fluorescence images, demonstrating high resolution and high accuracy 3D RI reconstruction of bulky and heterogeneous biological samples in vitro.
}




\maketitle

\section{Introduction}\label{sec1}

Fluorescence microscopy and phase microscopy are two powerful techniques in the field of biological imaging. Fluorescence microscopy captures molecular specified structural and functional information through exogenous fluorescence labeling or intrinsic autofluorescence. Phase microscopy quantitatively evaluates the biophysical properties of biological samples by measuring their refractive index (RI). Combining these distinct imaging modalities to image the same sample allows for studying the correlation between fluorescence-labeled structures and label-free phase structures with heterogeneous RI \cite{Xue:22, 2p-brief, Park:06, Kim:17, Chowdhury:17, Yeh:19, Dong2020-dq, shaffer2012single, marthy2024single, 9260959, tayal2020simultaneous, 9376596, Liu:18, pavillon2010cell,PHAM2021127290}. 

Most multimodal fluorescence-phase microscopy operates in transmission mode, capturing fluorescence images and phase images separately or sequentially \cite{Park:06, Kim:17, Chowdhury:17,  Yeh:19, shaffer2012single,  marthy2024single,9260959,tayal2020simultaneous, 9376596, Liu:18, pavillon2010cell, Dong2020-dq}. These two imaging modalities function independently, akin to separate microscopy systems, while sharing part of the optical paths. Phase images are captured using diffracted excitation light based on the configuration of optical/intensity diffraction tomography (ODT/IDT) \cite{Choi2007-ay, Sung:09, Waller2010-oj, Tian2015-ij}, while fluorescence images are formed from fluorophores excited by the same or different light sources. Although these techniques enable multimodal imaging of individual cells and even cellular organelles using objective lenses with high magnification and numerical aperture (NA) \cite{Park:06, Kim:17, Chowdhury:17,  Yeh:19, shaffer2012single,  marthy2024single,9260959,tayal2020simultaneous, 9376596, Liu:18, Dong2020-dq, pavillon2010cell}, the transmission mode limits their in vivo applications. This limitation arises mainly from the phase imaging modality, as back-scattered excitation light is much weaker than forward-scattered light. In contrast, epi-mode fluorescence microscopy is widely used because fluorescence emits isotropically in the $4 \pi$ space around the fluorophore. Reflective phase microscopy based on interferometry has also been developed \cite{Choi2014-rd, Kang2015-mz, Singh2019-zy, hyeon2021effect, Kang:23}, but none has demonstrated simultaneous imaging of fluorescence yet. 

Recently, our group \cite{Xue:22, 2p-brief} developed \textit{b}i-functional \textit{r}efractive \textit{i}nd\textit{e}x and \textit{f}luorescence microscopy (BRIEF) that \textit{experimentally} demonstrates simultaneous fluorescence and phase imaging in \textit{reflection mode} for the first time. BRIEF thoroughly merges these imaging modalities by reconstructing 3D RI from dozens of diffracted fluorescence images acquired in reflection mode. Each diffracted fluorescence image is illuminated by a single fluorophore embedded within the sample with the light being diffracted by phase objects located above the fluorophore. This process is modeled using a multi-slice model to describe light propagation through multiple scattering layers. The 3D RI of the phase objects is reconstructed by solving an inverse problem using gradient descent optimization. However, one-photon BRIEF requires contrived samples with sparsely distributed fluorescent beads to avoid crosstalk between measurements under one-photon excitation. Moreover, unlike the coherent or low-coherent plane wave used in ODT/IDT, BRIEF utilizes spatially varying, partially coherent spherical waves emitted by individual fluorophores within the sample. While these fluorescent sources enable optical sectioning, they also present challenges in precisely modeling the light field. In terms of computational efficiency, BRIEF encounters challenges similar to those of traditional ODT/IDT methods when processing large numbers of high-resolution forward images due to the computational burden.

To improve computational efficiency in phase recovery, many deep learning-based methods have been developed \cite{wang2024use, 10004797}. A classic approach is to train an end-to-end neural network, such as a convolutional neural network (CNN) \cite{Kamilov:15, Wu:22, Matlock:23, Zhou:20}, to directly retrieve the phase from input images. However, these end-to-end strategies typically require large, high-quality datasets and often suffer from low interpretability. 
Recently, the development of \textit{p}hysically \textit{i}nformed \textit{n}eural \textit{n}etworks (PINNs) \cite{Raissi2019-el, saba2022physics,Yang:23} and \textit{n}eural \textit{r}adiance \textit{f}ields (NeRFs) \cite{xu2022point, rzepecki2022fast} has offered new ways to solve ill-posed inverse problems \textit{without the need for large dataset}. The neural networks in NeRF-based method are treated as partial differential equation solvers, which significantly enhances the interpretability. This new paradigm has been widely adopted in various applications, such as ptychographic microscopy \cite{Zhou:23}, volumetric fluorescence imaging \cite{zhang2024single}, adaptive optics \cite{kang2024coordinate}, wavefront shaping \cite{Feng2023-ln}, computed tomography \cite{sun2021coil}, and dynamic structured illumination \cite{cao2022dynamic}. Unlike reconstructing light intensity using the ray optics rendering model in many NeRF-based methods, reconstructing the phase of light with the diffraction optics rendering model presents a more complex challenge. Liu et al.  \cite{liu2022recovery} combined neural fields with the Born approximation to solve the RI from discrete intensity-only measurements taken in transmission mode. However, due to the nature of implicit representations, which continuously encode unknown RI into the network, these networks struggle with strong nonlinearity or sharp gradient issues,  requiring long training time. Although phase retrieval using implicit NeRFs can successfully reconstruct RI, the use of overly deep neural networks and complex rendering equations significantly increases computational complexity, potentially making it difficult to achieve convergence to the global optimum. Conversely, shallow neural networks paired with simple rendering equations may be computational efficient, but lack sufficient constraints and representational capacity. Therefore, striking a balance between the complexity of the neural network and the rendering equation is crucial.


To overcome these limitations, we have developed fluorescence diffraction tomography (FDT) that uses explicit neural fields to reconstruct the 3D RI of label-free phase objects from diffracted two-photon fluorescence images in reflection mode. The explicit neural fields \cite{fridovich2022plenoxels} allow for faster 3D reconstruction compared to implicit neural fields \cite{zhang2020nerf++}. A differentiable multi-slice model for multiple scattering is used to render the forward diffraction. FDT also uses two-photon selective excitation \cite{Xue2019-fk} instead of one-photon excitation to remove the constraint of sparse fluorescence labeling. The key features of FDT are as follows:

    
    

\begin{enumerate}[label=\roman*)]
    \item We model the unknown RI using explicit neural representations and combine this with a coarse-to-fine structure to efficiently reconstruct high-resolution 3D RI.
    
    \item We develop a self-calibration method to accurately estimate fluorescent illumination on the phase objects.
    
    
    \item We design partially coherent masks to resolve the model discrepancy between the partially coherent light (i.e., fluorescence) used in the experiment and the coherent light assumed in the reconstruction process.
    
\end{enumerate}

We demonstrate FDT by reconstructing the 3D RI of bulky biological samples from diffracted fluorescence images collected in reflection mode. We successfully reconstruct the 3D RI of thin layers ($\sim$ 44 $\mu m$ thick) of living Madin-Darby Canine Kidney (MDCK) GII cells and a bulky 3D cultured bovine myotube ($\sim$ 300 $\mu m$ thick) using diffracted two-photon fluorescence images captured on a single $z$ plane. To our knowledge, FDT is the first diffraction tomography method using two-photon excited fluorescence images. FDT significantly advances fluorescence-phase multimodal imaging and potentially can be used for studying the interactions between fluorescence-labeled and label-free phase objects in bulky tissue.

\begin{figure}[t]
\includegraphics[width=\textwidth, trim={0 1120 0 0},clip]{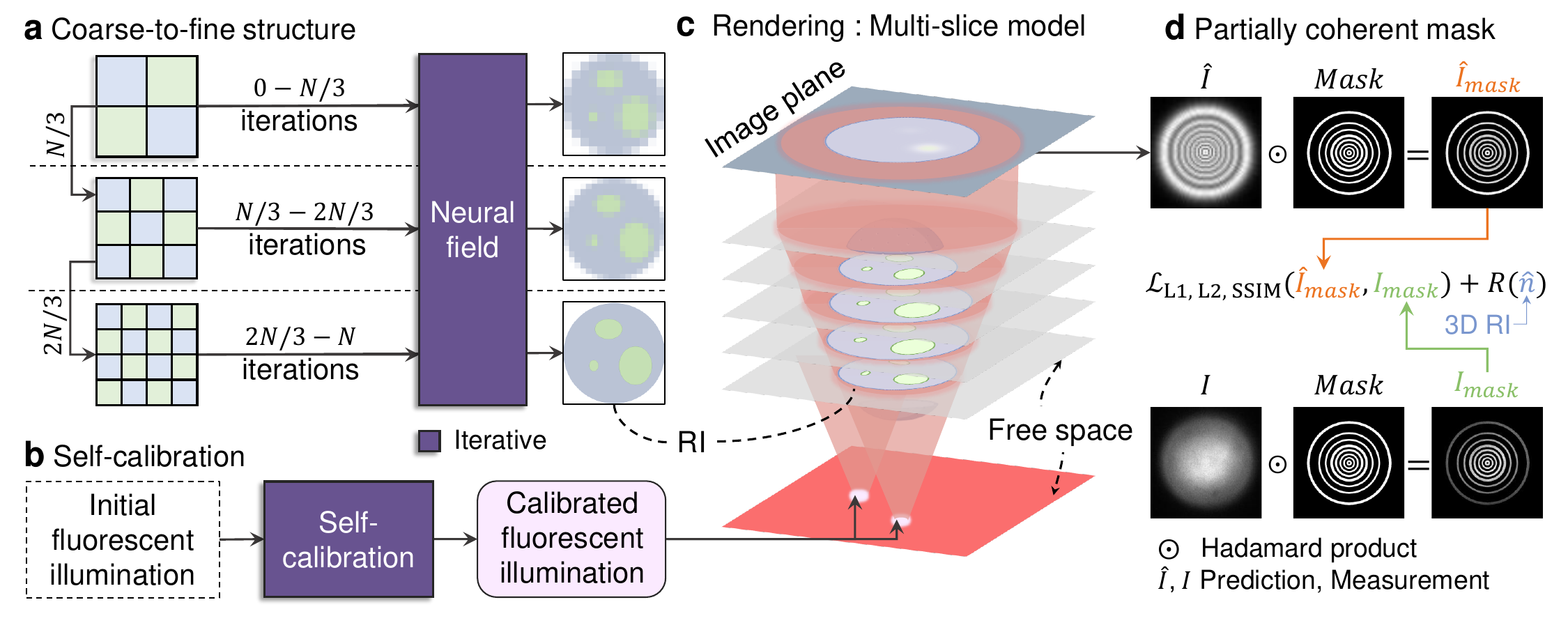}
\caption{\textbf{Overview of FDT using explicit neural fields.} \textbf{a}, The coarse-to-fine structure represents the unknown RI with neural fields and resolv it through three stages of increasing resolution as the number of iterations increases. \textbf{b}, Self-calibration adjusts the parameters to accurately model the fluorescent illumination for each measurement. These variables are then set as iterative parameters and optimized during the training process. \textbf{c}, The rendering equation is based on a differential multi-slice model, which takes two inputs: the RI from \textbf{a} and the fluorescent illumination from \textbf{b}. The model calculates the light field as it is modulated by the heterogeneous RI on each slice using the Born approximate. Fresnel propagation is used to calculate light propagation between slices.  \textbf{d}, A partially coherent light mask is applied to both the predicted and measured images to reduce the model mismatch. The masked images are used to calculate the loss function, incorporating L1, L2, SSIM, and regularization terms.}
\label{overview}
\end{figure}

\section{Results}\label{sec2}

\subsection{Model architecture}\label{subsec2.1}

The pipeline of the model architecture is shown in Fig. \ref{overview}. We first define 3D grids to explicitly represent the unknown 3D RI and set them as the parameters to be optimized within the PyTorch framework, as detailed in Section \ref{subsec3.2}. To reconstruct both low and high spatial frequency components, the 3D grids of the unknown 3D RI are trained using a three-stage coarse-to-fine structure (Fig. \ref{overview}a). The sampling resolution is gradually increased as the number of iterations increases. 

The reconstruction of RI using FDT relies on accurately modeling the fluorescent illumination at each point of the sample. Unlike standard ODT/IDT, which can calibrate the illumination field without imaging samples, in FDT, fluorescent illumination and label-free phase objects are intertwined and cannot be separated. The illumination field from a single excited fluorophore is spatially variant, as the excited fluorophore is a point source emitting spherical waves and close to the phase objects. Even though the positions of fluorescent sources are provided by the spatial light modulator in the optical system (Section \ref{subsubsec3.1.1}), we found that these values lack the precision needed to accurately reconstruct the 3D RI due to the mismatch between the partially coherent light in the experiment and the coherent light assumed in the model. To generate the forward images precisely, we have developed a self-calibration method (Fig. \ref{overview}b) that accurately calculates the illumination angles from each fluorescent source at every pixel. The illumination angle are determined by the original position of the fluorescent source, the lateral resolution of the sampling grid, the distance between the source and the phase objects (``free space"), the number of $z$-planes and the distance between adjacent $z$-planes. These variables are then set as iterative parameters and jointly optimized with the 3D RI. 



Next, both the neural representatives of the unknown 3D RI and the fluorescent illumination are input to a differentiable multi-slice model (Fig. \ref{overview}c) to generate the forward fluorescence images. The principle of the multi-slice model is the same as the one used in BRIEF, but we transfer it to a neural network framework for parallel forward image generation and jointly optimization of the 3D RI of phase objects and the self-calibration parameters. By leveraging the flexible training framework, we implement batch training and dynamically adjust the learning process to enhance the generalization and training speed. To simultaneously solve the 3D RI and perform self-calibration, we assign different learning rates to the parameters involved in self-calibration and activate the self-calibration process during the second coarse-to-fine stage to prioritize RI reconstruction.


To minimize the discrepancy between the partially coherent fluorescence and the coherent light used in the multi-slice model, we designed a partially coherent mask to filter both the predicted and measured image before calculating the loss (Fig. \ref{overview}d). Although partially coherent illumination can theoretically be modeled using the Wigner distribution function \cite{bastiaans1986application, Zuo2015-or}, it is impractical to apply this method to model fluorescence from each fluorophore due to the case-by-case variation in the degree of coherence. This variation is influenced by factors such as the type of fluorophores, the spatial focus of the two-photon excitation, and tissue scattering. Therefore, creating a partially coherent mask is a more feasible approach than using the Wigner function in our experiments. To generate the mask, we first generated reference images with a homogeneous RI of 1.33 (the background RI of cell culture solution) under coherent spherical illumination in the simulation, which is a defocused Airy pattern. We then binarize the defocused Airy pattern to select the bright areas as the partially coherent mask. The mask shown in Fig. \ref{overview}d is simplified for visualization purpose; in practice, the diffraction rings are much denser due to the large defocus distance (over 100 $\mu m$) and barely visible (Section \ref{subsec2.3}). 
Finally, we compute the Hadamard product of the partially coherent mask with the predicted image ($\hat{I}$) and the measured image ($I$), respectively, to generate the intensity-weighted predicted images, ($\hat{I}_{mask}$) and the intensity-weighted measured images (${I}_{mask}$). We compute $L1$, $L2$, and Structural Similarity Index Measure (SSIM) losses between $\hat{I}_{mask}$ and $I_{mask}$. Additionally, we apply a total variance (TV) regularizer to the predicted RI along the lateral axes ($R_{xy}$) and along the $z$ axis ($R_{z}$). Please see Section \ref{subsubsec3.1.2} for the detailed equations about the loss function. In summary, our model architecture integrates the coarse-to-fine structure, self-calibration, differentiable multi-slice neural network model, and partially coherent mask. This comprehensive approach ensures high fidelity in the forward fluorescence imaging process, achieving the high fidelity reconstruction of 3D RI.

\subsection{Model validation with simulated datasets}\label{subsec2.2}

%
\begin{figure}[htbp]
\includegraphics[width=\textwidth, trim={0 860 0 0},clip]{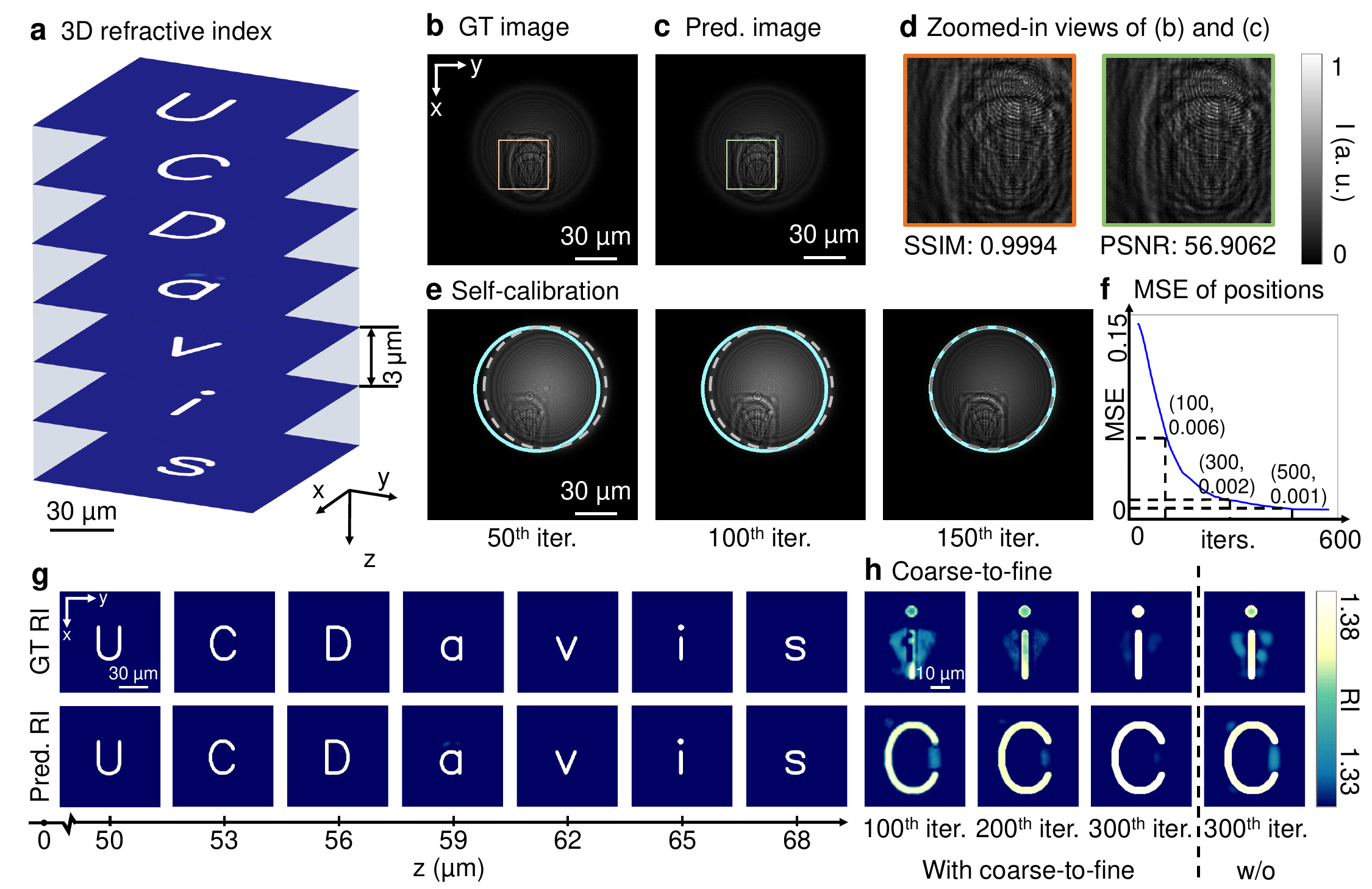}
\caption{\textbf{Reconstruction of the 3D RI of a simulated ``UCDavis" pattern.} \textbf{a,} Reconstructed 3D RI distribution of the ``UCDavis" pattern in 3D from 400 fluorescence images. \textbf{b,} An representative ground-truth (GT) image, generated using the multi-slice model with the ground-truth 3D RI, where the RI of the letters is 1.38 and background is 1.33. \textbf{c,} The predicted fluorescence image under the same illumination as \textbf{b} with the reconstructed 3D RI. \textbf{d,} Zoomed-in views of the regions within the orange and green boxes in \textbf{b} and \textbf{c}, respectively. The SSIM and PSNR between the ground-truth and predicted images are 0.9994 and 56.9062, respectively. \textbf{e,} Results of self-calibration of the positions of fluorescent sources at 50, 100, and 150 iterations. The blue circle indicates the irradiated region from the ground-truth position of the excited fluorophore, and the white dashed circle indicates the irradiated region from the predicted position of the fluorophore. \textbf{f,} The plot on the right shows the MSE loss between the self-calibrated and ground-truth fluorophore positions, converging to 0.001 within 500 iterations, indicating successful self-calibration of the positions of fluorescent sources. \textbf{g,} Comparison of the ground-truth RI (top row) and the predicted RI (bottom row) on each $z$-plane, as indicated by the axis below. \textbf{h,} Effect of the coarse-to-fine structure on reconstruction results on two different $z$-planes (top row, $z = 65$ $\mu m$; bottom row, $z = 53$ $\mu m$). The first three columns show results of the coarse-to-fine structure at different iterations (100, 200, and 300) with progressively increasing sampling grid at 128 $\times$ 128, 256 $\times$ 256, and 512 $\times$ 512 pixels. The last column shows the results after 300 iterations without the coarse-to-fine structure at a sampling grid of 512 $\times$ 512 pixels. Comparing the third column and the fourth column, the coarse-to-fine structure mitigates crosstalk between $z$-planes and reconstructs the missing low-frequency signals.}
\label{ucd}
\end{figure}

We first validate our method by reconstructing the 3D RI of a ``UCDavis" pattern with simulated data (Fig. \ref{ucd}). In the forward process, the ground-truth RI of the ``UCDavis" pattern consists of 14 layers separated by 3 $\mu$m, with a sampling grid of 512 $\times$ 512 pixels at a  resolution of 0.33 $\mu$m per pixel. Each letter in ``UCDavis" is located at the center of the odd layers, with homogeneous layers corresponding to the even layers. The RI value for the letters is 1.38 and the background RI is 1.33, which approximately matches the RI of biological cells and culture media \cite{Khan2021-vz}. We use the ground-truth RI (Fig. \ref{ucd}g, top row) to generate 400 ground-truth fluorescence images under 400 illuminations (one represented image is shown in Fig. \ref{ucd}b). All fluorescence images are captured on the same image plane, which corresponds to the top plane of the ``UCDavis" pattern. All the fluorophores are evenly distributed on a $z$-plane that is 60 $\mu m$ below the bottom $z$-layer of the ``UCDavis" pattern, forming a 20 $\times$ 20 grid in an 84.5 $\times$ 84.5 $\mu$m area, with the grid's center overlapping the center of the image. 


Our model is then trained on the 400 ground-truth fluorescence images to reconstruct the 3D RI, and then generate the predicted images (one represented predicted image is shown in Fig. \ref{ucd}c) based on the predicted RI (Fig. \ref{ucd}a, g) following the procedure described in Fig. \ref{overview} and Section \ref{subsec2.1}. 
Figure \ref{ucd}a shows the predicted RI of the 3D ``UCDavis" pattern, where each letter has a clear edge with no crosstalk between adjacent letters. This demonstrates that our model can successfully solve the RI of phase objects overlapping along the $z$-axis with excellent optical sectioning. The predicted forward image (Fig. \ref{ucd}c; zoomed-in view on the right in Fig. \ref{ucd}d) calculated using the reconstructed RI (Fig. \Ref{ucd}a) is very similar to the ground-truth forward image (Fig. \ref{ucd}b; zoomed-in view on the left in Fig. \ref{ucd}d), with a Mean Square Error (MSE) of $2.0388 \times 10^{-6}$, a SSIM of 0.9994, and a PSNR of 56.9062, indicating a successful reconstruction. The reconstructed RI of each letter closely matches the ground-truth RI (Fig. \ref{ucd}g). Further quantitative evaluation can be found in the supplementary, Table \hyperlink{appendix}{\ref*{tab:simulated_data2}}. These results demonstrate that our model is capable of accurately predicting the RI and forward images that closely resemble the ground truth. 

In the following subsection, we conduct an ablation study and validate each component of our method. In the comparison experiments, we did not compare our method with traditional optimization  \cite{Xue:22} or implicit neural networks  \cite{liu2022recovery} because these methods are based on entirely different optical setups without utilizing two-photon excited fluorescence as illumination sources and/or reconstructing 3D RI from fluorescent images. 

\subsubsection{Coarse-to-fine structure}\label{subsec2.2.1}

The coarse-to-fine structure divides the training process into three stages with a gradually increasing sampling grid in the $xy$ plane. In the simulation experiment described above (Fig. \ref{ucd}), the total number of training iterations is 300. During the first 100 iterations, a coarse sampling grid of 128 $\times$ 128 pixels is used to model the unknown RI. For the subsequent 100-200 iterations, the sampling grid is increased to 256 $\times$ 256 pixels. Finally, during the 200-300 iterations, the highest sampling grid of 512 $\times$ 512 pixels is used. The number of $z$-layers remains at 14 throughout all three stage. At the transition points, bilinear interpolation is employed to upsample the RI to ensure smooth transitions between different sampling grids. Both experiments are trained using the Adam optimizer with a learning rate of $5 \times 10^{-3}$, and each training run takes approximately 20 minutes for 300 iterations.

Figure \ref{ucd}h compares the reconstructed RI of two selected planes corresponding to the letters ``i" and ``C", which are 12 $\mu m$ apart along the $z$-axis, with and without using the coarse-to-fine structure. The first three columns illustrate the predicted RI of the letters ``i" and ``C" in ``UCDavis" at 100, 200, and 300 iterations, respectively. The sampling grids are gradually increased for both the RI and the predicted images from 128 $\times$ 128 to 256 $\times$ 256 and finally 512 $\times$ 512 pixels. The rightmost column shows the results after 300 iterations without employing the coarse-to-fine structure, which directly uses a sampling grid of 512 $\times$ 512 pixels in all iterations. Comparing these two cases, the coarse-to-fine structure reduces crosstalk in the RI map of ``i" and ``C", thereby enhancing the $z$-sectioning ability. Moreover, without the coarse-to-fine structure, directly solving high-resolution RI often leads to entrapment in local minima, thereby failing to accurately resolve the low spatial frequency of phase, such as the artificial dark center in the dot of ``i"  (Fig. \ref{ucd}h, 4\textsuperscript{th} column). In contrast, the coarse-to-fine structure successfully solve the low spatial frequency components, effectively eliminating the artificial dark center in the dot of ``i" (Fig. \ref{ucd}h, 3\textsuperscript{rd} column). The coarse-to-fine method improves the MSE of RI from \(1.8282 \times 10^{-6}\) to \(1.5730 \times 10^{-6}\) (Table \ref{tab:simulated_data2}). Figure \ref{ucd}h demonstrates that our coarse-to-fine structure effectively enhances the convergence and enables the accurate reconstruction of both low and high spatial frequency components, leading to superior quality reconstructions.

\subsubsection{Self-calibration of fluorescent illumination }\label{subsec2.2.2}

We next validate the self-calibration of fluorescent illumination by localizing fluorescent sources in the simulation in 3D (Fig. \ref{ucd}e-f). In the simulated data, the initial position $\boldsymbol{\hat{p}_i}=(\hat{p}_{ix}, \hat{p}_{iy}, \hat{p}_{iz})$ is randomly draw from a Gaussian distribution \(\mathcal{N}(0, \sigma^2)\) with uncertainty relative to the ground-truth position, defined as \(\boldsymbol{\hat{p}_i} = \boldsymbol{p_i} + \mathcal{N}(0, \sigma^2)\). The ground-truth position \(\boldsymbol{p_i}\) is normalized to the range $[-0.5, 0.5]$, while the standard deviation $\sigma$ is set to be 0.1 to simulate the case where the initial estimation of the position is significantly different from the ground-truth. In the experimental data (Section \ref{subsec2.3}), the initial position $\boldsymbol{\hat{p}_i}$ is determined by fitting a 2D Gaussian function to the defocused fluorescence images. The mean of the 2D Gaussian represents the lateral position of the fluorescent source, while the standard deviation corresponds to the defocus distance in the $z$-axis. For one defocused fluorescence image $I_i$, the initial fluorophore position $\boldsymbol{\hat{p}_i}$ is solved by:
\begin{equation}
\underset{\hat{p}_{ix}, \hat{p}_{iy}, \hat{p}_{iz}, A}{\operatorname{argmin}} \sum_{x=1}^{N} \sum_{y=1}^{N}  \left|\left|I_i(x,y) - A \exp\left\{  -\frac{1}{2\sigma_{z}(\hat{p}_{iz})^2} \left[(x - \hat{p}_{ix})^2+(y - \hat{p}_{iy})^2\right] \right\} \right|\right|_2 ^2. 
\end{equation}  
These positions are then set as iterative parameters that are updated continuously throughout the training process. 

Figure \ref{ucd}e shows the intermediate results of self-calibration at 50, 100, and 150 iterations with the simulated data ``UCDavis". The illuminated area in the predicted fluorescence image (dash white circle) gradually aligns with the fluorescent illumination from the ground-truth position (blue solid circle), indicating that the estimated position of the fluorescent source is converging towards the ground-truth position. 
Quantitatively, the MSE of the positions decreases rapidly as the number of iterations increases (Fig. \ref{ucd}f),  converging to a lower error value in 20 minutes. The self-calibration method significantly improves the SSIM of the images from 0.9723 to 0.9933 and reduces the MSE of RI from $2.7980 \times 10^{-5}$ to $3.1984 \times 10^{-6}$ (Table \ref{tab:simulated_data2}, and Fig. \hyperlink{appendix}{\ref*{ucd_sup}} within 300 iterations. Besides calibrating the positions of fluorescent sources, the self-calibration also adjusts all parameters related to modeling the illumination, including the lateral resolution of the sampling grid, the distance between the source and the phase objects (``free space"), the number of $z$-planes, and the distance between adjacent $z$-planes. Quantitative comparisons with and without self-calibration of these parameters are presented in Fig. \hyperlink{appendix}{\ref*{ucd_sup}} and Table \hyperlink{appendix}{\ref*{tab:simulated_data2}}. In summary, the self-calibration method accurately and efficiently estimates the positions of fluorescent sources, as an essential step for reconstruction RI using FDT.

\subsubsection{Differential multi-slice model for 3D rendering}\label{subsec2.2.3}

To improve computational efficiency and flexibility, we extend the conventional multi-slice model used in BRIEF \cite{Xue:22} to the PyTorch framework. The differentiable multi-slice model with automated backpropagation mechanisms allows us to adjust the model's parameters arbitrarily. This flexibility enables fine-tuning of the RI resolution using the coarse-to-fine strategy and optimizing the self-calibration parameters for fluorescent illumination, as discussed earlier. In the multi-slice model, the 3D phase object is modeled as a stack of multiple $z$-layers, each with an unknown RI $\hat{n}_k(\mathbf{r})$, where $k = 1, 2, ..., N_z$. The 3D RI of a total of $N_z$ layers is:

\begin{equation}
\hat{n} \triangleq \left\{ \hat{n}_k(\mathbf{r}) \right\}_{k=1}^{N_z}, \quad \mathbf{r} = (x, y),
\end{equation}
where $\mathbf{r} = (x, y)$ is the spatial coordinates. As light propagates through each layer, its phase is altered according to the transmission function \( t_k(\mathbf{r}) \) at $k^{th}$ layer: 
\begin{equation}
t_k(\mathbf{r}) = \exp \left( \frac{j 2\pi}{\lambda} \Delta z (\hat{n}_k(\mathbf{r}) - n_b) \right), \quad k=1, 2, \ldots, N_z,
\label{first}
\end{equation}
where $\lambda$ is the wavelength of the light, $\Delta z$ is the thickness of each layer, and $n_b$ is the RI of the background medium. We then use the operator $\mathcal{P}_{\Delta z}$ to represent the Fresnel propagation of the field over a distance $\Delta z$:
\begin{equation}
\mathcal{P}_{\Delta z} \{ \cdot \} = \mathcal{F}^{-1} \left\{ \exp\left(-j 2\pi \Delta z \sqrt{\left(\frac{1}{\lambda}\right)^2 - ||\mathbf{k}||^2} \right) \cdot \mathcal{F} \{ \cdot \} \right\}, \quad \mathbf{k} = (k_x, k_y),
\end{equation}
where $\mathcal{F} \{ \cdot \}$ and $\mathcal{F}^{-1} \{ \cdot \}$ denote the Fourier transform and its inverse, respectively, and $\mathbf{k} = (k_x, k_y)$ is the spatial frequency coordinates. The electric field from the $i^{th}$ fluorescent source after propagating through the $k^{th}$ layer is described by:
\begin{equation}
\hat{E}_{k,i}(\mathbf{r}) = \mathcal{P}_{\Delta z} \left\{ t_k(\mathbf{r}) \cdot \hat{E}_{k-1,i}(\mathbf{r}) \right\},
\end{equation}
where $\hat{E}_{k-1,i}(\mathbf{r})$ is the electric field at the $(k-1)^{th}$ layer and $\hat{E}_{k,i}(\mathbf{r})$ is the electric field at the $k^{th}$ layer. If the image plane is at the top surface of the phase object (the final $z$-layer), the electric field at the camera under $i^{th}$ fluorescent illumination can be calculated by applying the pupil function $p(\mathbf{k})$ to the field at the final layer:
\begin{equation}
\hat{E}_i(\mathbf{r}) = \mathcal{F}^{-1} \left\{ p(\mathbf{k}) \cdot \mathcal{F} \left\{ \hat{E}_{N_z,i}(\mathbf{r}) \right\} \right\},
\end{equation}
where $\hat{E}_{N_z,i}(\mathbf{r})$ is the electric field at the final layer. If the image plane is $\Delta Z_c$ below the final layer, as is the case when the phase object is thick (Section \ref{subsec2.3.2}), the field is first back-propagated over a distance $\Delta Z_c$ before passing through the pupil and arriving at the camera:
\begin{equation}
\hat{E}_i(\mathbf{r}) = \mathcal{F}^{-1} \left\{ p(\mathbf{k}) \cdot \mathcal{F} \left\{ \mathcal{P}_{-\Delta Z_c} \left\{\hat{E}_{N_z,i}(\mathbf{r}) \right\} \right\}\right\},
\label{b2c}
\end{equation}
where $\mathcal{P}_{-\Delta Z_c}$ is the back-propagation operator. Since the camera can only detect the intensity of light field, the intensity image captured under $i^{th}$ fluorescent illumination is described as:
\begin{equation}
\hat{I}_i(\mathbf{r}) = | \hat{E}_i(\mathbf{r}) |^2.
\end{equation}

This mathematical model enables the simulation of the electric field's propagation through the layers of a sample within the PyTorch framework. However, the model assumes coherent light, while in exepriments, two-photon excited fluorescence is partially coherent light sources and spatially varying (depending on the local fluorescent label), leading to a mismatch between the model and actual experiments. We address this issue using a partially coherent mask as demonstrated in the next section with experimental data.

\subsection{Evaluation with experimental data of biological samples}\label{subsec2.3}

In this section, we evaluate FDT by reconstructing the 3D RI of real biological samples, including 2D cultured living MDCK cells and 3D cultured bovine muscle stem cells forming a bulky bovine myotube. The label-free cells are cultured on a coverslip-bottom dish with fluorescent dye sprayed on the outside. Detailed information on sample preparation and the optical setup is provided in Section \ref{subsec3.1}.

\subsubsection{3D RI reconstruction and analysis of MDCK cells}\label{subsec2.3.1}
\begin{figure}[htbp!]
\includegraphics[width=\textwidth, trim={0 580 0 0},clip]{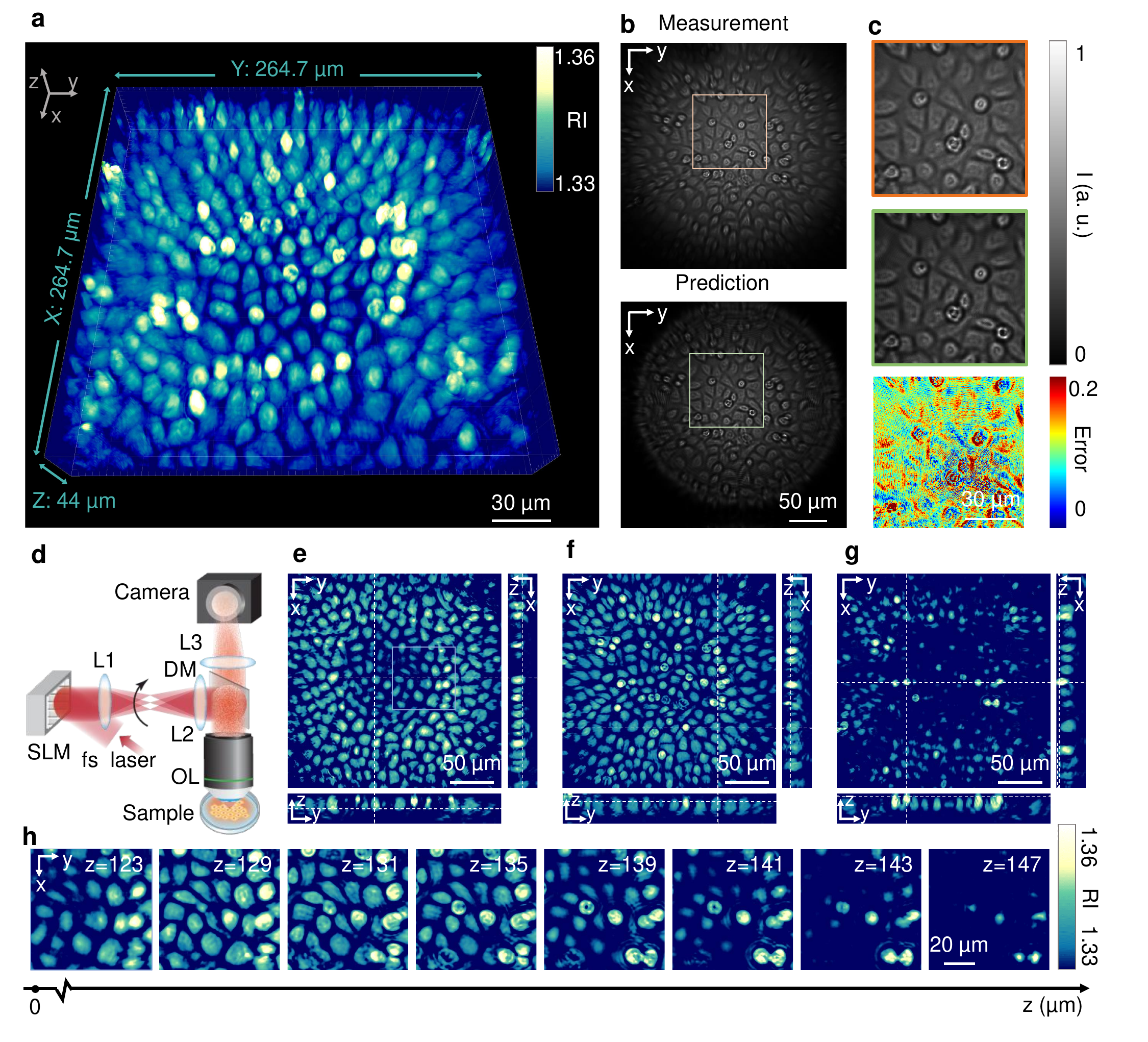}
\caption{\textbf{3D RI reconstruction of a thin layer of live MDCK Cells sample.} \textbf{a}, 3D visualization of the RI distribution of MDCK cells within a 358.4 $\times$ 358.4 $\times$ 44 $\mu m^3$ volume. The RI of the cells ranges between 1.33 to 1.36.  \textbf{b}, Comparison of the measured (top) and predicted (bottom) images. \textbf{c}, Zoomed-in views of the highlighted region: the measured image (top), the predicted image (middle), and the error map between the measured and predicted images (bottom). \textbf{d}, Schematic diagram of the optical setup of FDT. Fluorescence is excited by scanning a focus with a spatial light modulator (SLM), and diffracted fluorescence images are captured in reflection mode using a camera. \textbf{e-g}, Cross-sectional views of the RI distribution of MDCK cells on three representative planes that are 12.5 $\mu m$ apart, showing optical sectioning ability and high 3D resolution. \textbf{h}, Zoomed-in view of the image $z$-stack of cells in the highlighted region in (e). The $z$-position of each image is labeled on the $z$ axis below the images. The images again highlight the optical sectioning and high resolution of FDT. See the Supplementary Video for a better 3D visualization.}

\label{mdck1}
\end{figure}

We first experimentally validate FDT using a thin layer of MDCK cells (Fig. \ref{mdck1}). We excite 21 $\times$ 21 fluorescent spots one-by-one within a 167 $\times$ 169 $\mu m^2$ region on a single plane, located 105 $\mu m$ below the MDCK cells (measured by self-calibration). All fluorescent images are taken on a single image plane slightly above the MDCK cells. A total of 390 fluorescence images are selected for reconstructing the 3D RI of the MDCK cells. The reconstructed RI (Fig. \ref{mdck1}a) consists of 22 slices with a sampling grid of 1024 $\times$ 1024 pixels, forming a volume of $ 358.4 \times 358.4 \times 44 ~ \mu m^3 $. The reconstructed RI ranges from 1.33 to 1.36, demonstrating FDT is capable of quantitatively reconstructing 3D RI with high sensitivity and over a large field-of-view. The FDT method not only quantitatively reconstructs the spatial distribution of RI, but also demonstrates high accuracy in the predicted forward images, as seen in Fig \ref{mdck1}b. The zoomed-in view (Fig. \ref{mdck1}c) shows the predicted image (middle) successfully reconstruct the distinctive intracellular structures of the cells, as shown in the measured image (top). The pixel-wise error map (Fig. \ref{mdck1}c, bottom) shows that the differences between the measured and predicted images are minimal. In the predicted image, regions where cells have detached from the substrate show relatively larger errors, while regions where cells remain adherent to the substrate are more accurate. These accurate forward images are fundamental to the successful RI reconstruction. Figure \ref{mdck1}d shows the optical schematic diagram of FDT (details in Section \ref{subsubsec3.1.1}). The representative cross-sectional views of the RI in the $xy$, $xz$, and $yz$ planes (Fig. \ref{mdck1}e-g) reveal individual cells clearly, highlighting FDT's ability to achieve high resolution in 3D and excellent $z$-sectioning ability. In the zoomed-in view of the image stack along the $z$ axis (Fig. \ref{mdck1}h), cells can be observed appearing and disappearing. Most cells are alive and attached to the bottom of the dish (Fig. \ref{mdck1}e-f), while in the top several layers (Fig. \ref{mdck1}g), dead cells shrink, disassemble their focal adhesions, and float above the substrate. The relatively higher RI of dead cells compared to live cells probably indicates changes in intracellular organelles during cell death, such as the condense of chromatin. This result experimentally and quantitatively evaluates our method's $z$-sectioning capability, high resolution, and high sensitivity.


During the reconstruction process, partially coherent masks are applied to mitigate the mismatch between experimental data and the simulation model. To evaluate their effectiveness, we quantitatively compare the experimentally measured images with the predicted images, both with and without the masks (Fig. \hyperlink{appendixfigS2}{\ref*{mdck_sup}}, Table \ref{mdck_mask}). Although both approaches enable the reconstruction of the 3D RI of MDCK cells, the reconstruction using partially coherent masks is more accurate, as indicated by the comparison between the predicted and measured images (Fig. \hyperlink{appendix}{\ref*{mdck_sup}}). Quantitatively, without the masks, the mismatch in the fluorescence illumination field due to light coherence results in a relatively inaccurate reconstruction, with an MSE of $1.3453 \times 10^{-3}$. After applying the masks in the training process, the model mismatch is reduced, leading to a decrease in MSE to $6.2900 \times 10^{-4}$. Besides pixel-wise MSE, we also quantitatively compared other metrics, such as structural similarity (Table \ref{mdck_mask}). The SSIM improves significantly after applying the mask, increasing from 0.8177 to 0.9160. In addition, despite the predicted image being synthesized under coherent illumination and the measured image  under partially coherent illumination (Fig. \ref{mdck1}b-c), the high accuracy of the RI reconstruction after applying the partially coherent masks makes these two images very similar. In conclusion, partially coherent masks provide an effective and efficient approach to handling partially coherent illumination in FDT. 

\begin{table}[hbt!]
\centering
\caption{Performance metrics with and without the  partially coherent mask for MDCK cell reconstruction.}
\begin{tabular}{lcccc}
\hline
\textbf{Method}  & \textbf{MSE} & \textbf{SSIM} & \textbf{LPIPS} & \textbf{PSNR}  \\
\hline
w/  mask & 6.2900 $\times 10^{-4}$ & 0.9160 & 0.1417 & 32.0135  \\
w/o mask  & 1.3453 $\times 10^{-3}$ & 0.8177 & 0.1776 & 28.7119  \\
\hline
\end{tabular}
\label{mdck_mask}
\end{table}

\subsubsection{3D RI reconstruction and analysis of a bulky 3D cultured bovine myotube}\label{subsec2.3.2}

\begin{figure}[htbp]
\includegraphics[width=\textwidth, trim={0 630 0 0},clip]{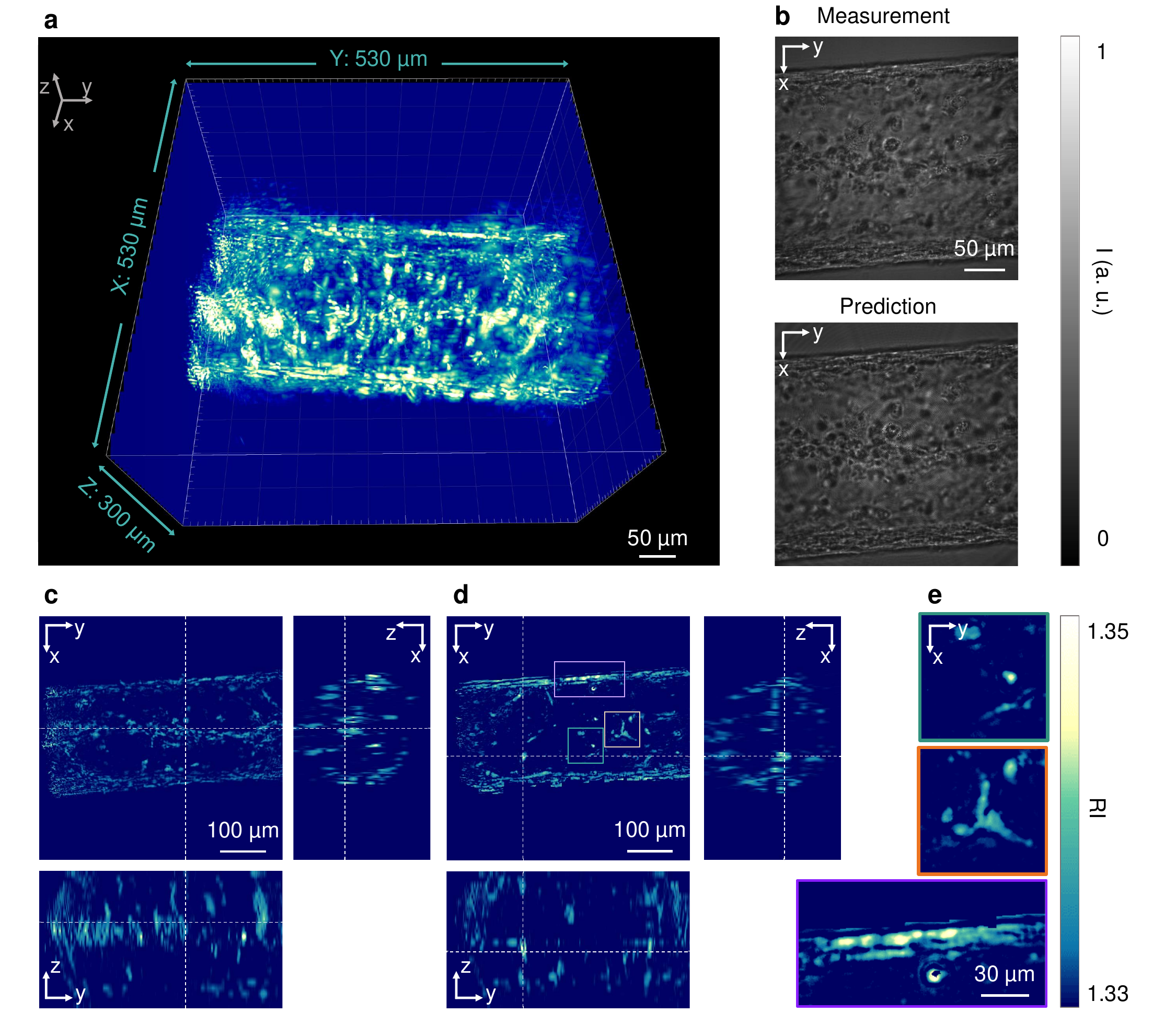}
\caption{\textbf{3D RI reconstruction of a 3D cultured bovine myotube.} \textbf{a}, 3D visualization of the RI of 3D cultured bovine myotube within a volume of $530 \times 530 \times 300 ~\mu m^3$. \textbf{b}, Comparison between the measured (top) and predicted (bottom) images of a representative region. \textbf{c-d}, Cross-sectional views of the reconstructed RI on two representative planes, showing high resolution and optical sectioning ability. \textbf{e}, Zoomed-in details of the highlighted regions in \textbf{d}, labeled by corresponding colors, showing different morphologies of stem cells in the 3D cultured bovine myotube during proliferation and differentiation. The results indicate that FDT can accurately reconstruct various structures across a wide range of spatial frequencies. See the Supplementary Video for a better 3D visualization.}
\label{3dtube}
\end{figure}

We next evaluate FDT by reconstructing the RI of a 3D cultured bovine myotube with complex structures (Fig. \ref{3dtube}). Unlike the thin layer of cells discussed in the previous section, the bovine myotube is significantly thicker (about 300 $\mu m$, Fig. \ref{3dtube}a) and consists of multiple layers of muscle cells. These muscle cells are differentiated from label-free muscle stem cells cultured in 3D. The myotube was fixed and placed on a Petri dish coated with fluorescent paint on the outer surface. A total of 21 $\times$ 21 fluorescent spots are excited one-by-one within a 167 $\times$ 169 $\mu m^2$ region on a single plane, located 210 $\mu m$ below the myotube. 
Given that the bovine myotube is much thicker and more complex than the MDCK cells, the image plane is positioned 90 $\mu m$ below the top surface of the myotube to optimize the SNR of forward images rather than at the top of the sample. Correspondingly, the reconstruction process implements the back-propagation operation (Eq. \ref{b2c}) as well. After assessing the SNR of images, 348 fluorescence images are selected for 3D RI reconstruction in a $530 \times 530 \times 300 ~\mu m^3$ volume, consisting of 24 $z$-slices with a sampling grid of $1024 \times 1024$ pixels. Individual cells are not distinguishable in the scrambled forward images (Fig. \ref{3dtube}b, top), but appear as bright and dark blurry shadows in the image depending on if they are above or below the image plane. The predicted image (Fig. \ref{3dtube}b, bottom) calculated with the reconstructed 3D RI closely matches the experimentally measured image, achieving a low MSE of $2.8607 \times 10^{-3}$. 
Cross-sectional views (Fig. \ref{3dtube}c-d) show the reconstructed RI of the 3D cultured bovine myotube on two representative $z$-planes that are 50 $\mu m$ apart in the entire volume. These orthogonal views clearly reveal the tubular morphology of the sample. The circular cross-section in the $xz$ view confirms the integrity of the tube structure, while the clear delineation between the top and bottom of the tube highlights the effectiveness of our method's $z$-sectioning ability. The distinctive distribution of cells inside the tube on the two different $z$-planes also show the excellent $z$-sectioning ability of FDT. The reconstruction also achieves high 3D resolution. Individual cells are visible and distinguishable after reconstruction, whereas they are indistinguishable in the forward images. The reconstruction is sensitive to RI differences of less than 0.02, highlighting FDT's high sensitivity for quantitative phase imaging. 



Our results reveal individual cells with diverse morphological structures and RI during the processes of differentiation and aggregation (Fig. \ref{3dtube}e). This demonstrates that FDT not only achieves high resolution but also effectively reconstructs structures across both low and high spatial frequencies. The morphology of the muscle cells within the myotube can be categorized into three distinct differentiation stages: individual stem cells exhibiting membrane spreading (Fig. \ref{3dtube}e, top; Fig. \ref{3dtube}d, green box), interconnected cells forming pre-differentiation clusters within the matrix (Fig. \ref{3dtube}e, middle; Fig. \ref{3dtube}d, orange box), and elongated, stable muscle cells forming axial connections (Fig. \ref{3dtube}e, bottom; Fig. \ref{3dtube}d, purple box). The individual stem cells and the stable muscle cells exhibit a relatively higher RI compared to the interconnected cells, demonstrating RI as a potential biomarker to indicate the differentiation status of stem cells. Therefore, our method not only resolves cellular structures at high resolution but also quantitatively measures RI changes in cells at various metabolic states.




\subsection{Discussion}\label{subsec3}
In this study, we develop and validate FDT for reconstructing the 3D RI of label-free biological samples from epi-mode fluorescence images using explicit neural fields. FDT incorporates several innovations to accurately and efficiently reconstruct 3D RI with high spatial resolution and excellent optical sectioning ability. We apply a coarse-to-fine structure to represent the unknown RI for reconstructing both low and high spatial frequency components. To model the illumination sources, we accurately estimate the positions of excited fluorophores and the parameters of the sampling grids by self-calibration and  apply partially coherent masks to mitigate the model mismatching caused by partially coherent illumination. Finally, we upgrade the multi-slice model that calculates light diffraction and propagation in bulky tissue to pyTorch framework, largely improving the computing efficiency. 

To validate FDT, we first reconstruct the 3D RI of simulated data, specifically the seven letters in ``UCDavis" stacked in 3D. The results quantitatively prove that FDT has excellent optical sectioning ability and board spatial frequency coverage. We next reconstruct the 3D RI of biological samples using experimentally captured fluorescent images. We reconstruct the 3D RI of MDCK cells in a $358.4 \times 358.4 \times 44$ $\mu m^3$ volume, demonstrating the method's optical sectioning ability, high spatial resolution, and the effectiveness of the partially coherent mask. To demonstrate our method with more complex and bulky tissue, we reconstruct the 3D cultured bovine myotube in a $530 \times 530 \times 300$ $\mu m^3$ volume. We successfully resolve the tube structure, identify the RI of individual cells, and detects changes in the RI of cells at various metabolic states. Our method computes the RI of 300 $\mu m$-thick myotube ($1024 \times 1024 \times 24$ sampling grid) at subcellular resolution within one hour, achieving 90\% of the best result in just 20 minutes. This demonstrates that our approach is more robust, more accurate, and more computationally efficient than current state-of-the-art methods.

Among the innovations of FDT, the coarse-to-fine strategy significantly enhances both the speed and resolution of 3D RI reconstruction. This approach makes our method faster, more interpretable, and adaptable for future advancements, such as the octree structure \cite{yu2021plenoctrees}. Furthermore, the differential multi-slice model ensures that each variable is retrievable, which enables the self-calibration method, improves computational efficiency, and simplifies the code architecture. The multi-slice model and self-calibration will potentially be widely adopted in various methods for 3D rendering from multiple views.

Compared to our previous work BRIEF, FDT can handle dense objects by leveraging two-photon excitation instead of one-photon excitation, removing the constrain on fluorescence labeling and opening the avenue for boarder and more practice biological applications. By representing the 3D RI with explicit neural fields, FDT can reconstruct the RI of much more complex phase objects over a larger volume and in less time compared to BRIEF.  

Compared to state-of-the-art IDT using neural fields for RI reconstruction, such as DeCAF \cite{liu2022recovery}, our method uses explicit neural fields rather than implicit neural fields, which largely accelerates reconstruction speed. Our method also uses the multi-slice model for multiple scattering instead of the one-time first Born approximation for 3D rendering, allowing us to reconstruct the RI of the 300 $\mu m$-thick myotube. Unlike conventional ODT methods that use transmitted light and/or interferometry, our method is based on fluorescence microscopy in reflection mode, opening up the possibility of in vivo imaging in the future.

\textit{Limitations of FDT.} Although our method have successfully achieved diffraction tomography using two-photon excited fluorescence for the first time, there are still several aspects to be further optimized for more practical applications to be further optimized for more practical applications. First, our current setup includes a spacer with homogeneous RI (i.e., a coverslip) between the fluorescent sources and label-free biological samples, which could probably be removed after upgrading our model both computationally and optically. Second, our model currently faces a memory inefficiency issue, where the number of layers that can be processed is limited by the available GPU memory. This issue could potentially be resolved by developing memory-efficient methods to enhance the scalability of our approach. Addressing these limitations could improve the practicality and effectiveness of FDT for more complex and realistic biological imaging scenarios in the future.

\subsection{Conclusion}
Our FDT method significantly advances diffraction tomography with fluorescence illumination through explicit neural representations, achieving high-speed, high-resolution, and high-accuracy reconstruction of 3D RI. 
Unlike diffraction tomography in transmission mode, FDT in reflection mode enables multimodal imaging of bulky samples, and potentially paves the way for in vivo multimodal imaging. FDT has been demonstrated to quantitatively evaluate the RI changes of stem cells at various metabolic states and potentially could be applied to understanding fundamental biological processes of stem cells, as well as facilitate the development of medical therapies for degenerative diseases. 
\section{Method}\label{sec3}

\subsection{Experimental setup}\label{subsec3.1}

\subsubsection{Optical setup}\label{subsubsec3.1.1}
The laser source for FDT is a femtosecond laser at 1035 nm wavelength and 1 MHz repetition rate (Monaco 1035-40-40 LX, Coherent NA Inc., U.S.). A polarizing beam splitter cube (PBS123, Thorlabs) and a half-wave-plate (WPHSM05-1310) mounted on a rotation mount (PRM05, Thorlabs) are used to manually adjust the input power to the following optics in the system. The laser beam is collimated and expanded by a 4-f system (L1, LA1401-B, $f_1$ = 60 mm, Thorlabs; L2, LA1979-B, $f_2$ = 200 mm, Thorlabs) before reaching a SLM (HSPDM1K-900-1200-PC8, Meadowlark Optics Inc., U.S.) placed on the Fourier plane. The SLM modulates the phase of the beam to selectively excite fluorophore at any given position in 3D. After the SLM, the laser beam is relayed by two 4-f systems (L3, LB-1889-C, $f_3$ =250 mm, Thorlabs; L4, AC508-250-C-ML, $f_4$ = 250 mm, Thorlabs; L5, AC508-200-C-ML, $f_5$ =200 mm, Thorlabs; L6, AC508-200-C-ML, $f_6$ = 200 mm, Thorlabs). Additionally, a black spot is painted on a mirror placed at the back focal plane of L3 to block the zero-order beam after the SLM. Next, the modulated laser beam is reflected by a dichroic mirror (FF880-SDI01-T3-35X52, Chroma, U.S.) to the back-aperture of the objective lens (XLUMPLFLN20XW, 20$\times$, NA 1.0, Olympus Inc., Japan). Samples are placed on a manual 3-axis translation stage (MDT616, Thorlabs). In the emission path, a shortpass filter (ET750sp-2p8, CHROMA) is used to block the reflected excitation light. A bandpass filter (AT635/60m, CHROMA) is placed before the camera to pass through red fluorescence. The defocused fluorescence images are captured by a camera (Kinetix22, Teledyne Photometrics, U.S.) controlled by the Micro-Manager software. In addition, a one-photon widefield microscope is implemented, merged with the FDT setup, to locate the sample and find the focal plane before two-photon imaging. The one-photon system consists of an LED (M565L3, Thorlabs), an aspherical condenser lens (ACL25416U-A) to collimate the LED light, and a dichroic mirror (AT600dc, CHROMA) to combine the one-photon path to the two-photon path. The FDT system is controlled by a computer (OptiPlex 5000 Tower, Dell) using MATLAB (MathWorks, U.S.) and a data acquisition card (PCIe-6363, X series DAQ, National Instruments) for signal input/output. Customized MATLAB code is used to generate digital signals, and the NI-DAQ card outputs the signals to synchronously trigger the laser, the SLM, and the camera. For the experiment in Section \ref{subsec2.3.1}, the exposure time of each fluorescence image is 20 ms, and the total imaging time to collect 441 fluorescence images is 8.82 s. For the experiment in Section \ref{subsec2.3.2}, the exposure time of each fluorescence image is 100 ms, and the total imaging time to collect 441 fluorescence images is 44.1 s. 

\subsubsection{Computational Setting}\label{subsubsec3.1.2}
Our model is trained on an NVIDIA A6000 GPU with 48 GB of Memory. We use the Adam optimizer with an initial learning rate of $5 \times 10^{-3}$, a momentum decay rate of 0.9, and a decay rate of 0.99 for squared gradients. The batch size is set to 30. 

\textbf{Loss function.} The loss function is designed to optimize multiple aspects of the model's performance by incorporating various components that address distinct error metrics and regularization terms. The formulation of the loss function is as follows:
\begin{equation}
\mathcal{L}(\hat{n}; I_1, I_2, \ldots, I_n) = \sum_{i=1}^{N_B}\mathcal{L}(\hat{n}; I_i),
\end{equation}
where $\hat{n}$ is the predicted RI, $I_i$ represents the $i^{th}$ measurement of diffracted fluorescent image, and $N_B$ is batch size. The loss between each measurement and prediction is calculated as: 
\begin{equation}
\mathcal{L}(\hat{n}; I_i) = \alpha \|\hat{I_i} - I_i\|_1 + \beta \|\hat{I_i} - I_i\|_2^2 + \gamma \mathscr{\ell_i}_{\text{ssim}}(\hat{n}; I_i) + \tau_{xy} R_{xy}(\hat{n}) + \tau_z R_z(\hat{n}),
\end{equation}
where $\hat{I}$ denotes the predicted images generated by our multi-slice model. Each term in the loss function is weighted by specific parameters ($\alpha$, $\beta$, $\gamma$, $\tau_{xy}$, and $\tau_z$) balance their contributions based on the model's objectives. These hyperparameters are carefully tuned to optimize performance for the specific application. The L1 loss, representing the mean absolute error, promotes sparsity in predictions, while the L2 loss, representing the mean squared error, helps minimize large deviations. Initially, $\alpha$ is set to 0 to focus on reducing large errors, and it is gradually increased during training to shift the emphasis towards sparsity and fine-tuning with $\beta$ decreases correspondingly, where $\alpha + \beta = \text{constant}.$ This dynamic adjustment allows the model to start by making broad corrections and then progressively refine its outputs, creating a robust framework for achieving high-quality predictions. Additionally, the SSIM loss is included to assess the similarity between predicted and ground-truth images, capturing perceptual differences and preserving image quality, as described below: 
\begin{equation}
\mathscr{\ell}_{\text{ssim}}(\hat{n}; I_i) = 1 - \text{SSIM}(\hat{I_i}, I_i),
~~
\text{SSIM}(x, y) = \frac{(2\mu_x\mu_y + c_1)(2\sigma_{xy} + c_2)}{(\mu_x^2 + \mu_y^2 + c_1)(\sigma_x^2 + \sigma_y^2 + c_2)}.
\end{equation}
Furthermore, to promote smoothness in the image along the $x$, $y$, and $z$ axes, we incorporate Total Variation (TV) regularization terms. Given the differing resolutions in the $xy$ and $z$ planes, we employ two separate weighting terms to control the contribution of these regularizers, which are defined as:
\begin{equation}
R_{xy}(\hat{n}) = \|\nabla_{x} \hat{n}\|_1 + \|\nabla_{y} \hat{n}\|_1,
~~
R_z(\hat{n}) = \|\nabla_z \hat{n}\|_1.
\end{equation}

\textbf{Evaluation metrics.} The evaluation metrics used in this study include MSE for measuring the average squared differences between predicted and actual values; SSIM for assessing perceptual similarity by considering luminance, contrast, and structure; Learned Perceptual Image Patch Similarity (LPIPS) for evaluating perceptual similarity based on deep features; and Peak Signal-to-Noise Ratio (PSNR) for indicating the peak error between images. These metrics collectively provide a comprehensive evaluation of the model's predictions in terms of both numerical accuracy and perceptual quality.

\subsubsection{Sample preparation.}\label{subsubsec3.1.3}


\textbf{MDCK cells}. MDCK (Madin-Darby Canine Kidney) GII cells were cultured at $37^{\circ}$, 5$\%$ CO\textsubscript{2} in DMEM (Gibco) supplemented with 10$\%$ fetal bovine serum (FBS – RD Biosciences) and antibiotics. The cell line was checked for mycoplasma and routinely treated with mycoplasma removal agent (MRA) for preventive maintenance. The cells were plated onto a glass bottom dish (CellVis) pre-coated with rat tail collagen to promote cell adhesion to the substrate. 

\textbf{3D cultured bovine myotube.} Bovine muscle stem cells were isolated from freshly slaughtered Angus cow semitendinosus muscle received from the UC Davis Meat Lab by adapting a previously reported protocol \cite{mdck_paper, brassard2021recapitulating}. Myotubes differentiated from the stem cells were fabricated employing an advanced Matrigel-agarose core-shell bioprinting technique. The filaments were subsequently cultured in Ham’s F10 medium, enriched with 2$\%$ fetal bovine serum (FBS) and 1$\%$ penicillin-streptomycin (P/S). After 14 days, the well-differentiated bovine myotubes were carefully harvested from the filaments by dismantling the agarose shell. 

\subsection{Explicit representations used in FDT}\label{subsec3.2}
  
\begin{figure}[htbp]
\includegraphics[width=\textwidth, trim={0 1200 0 0},clip]{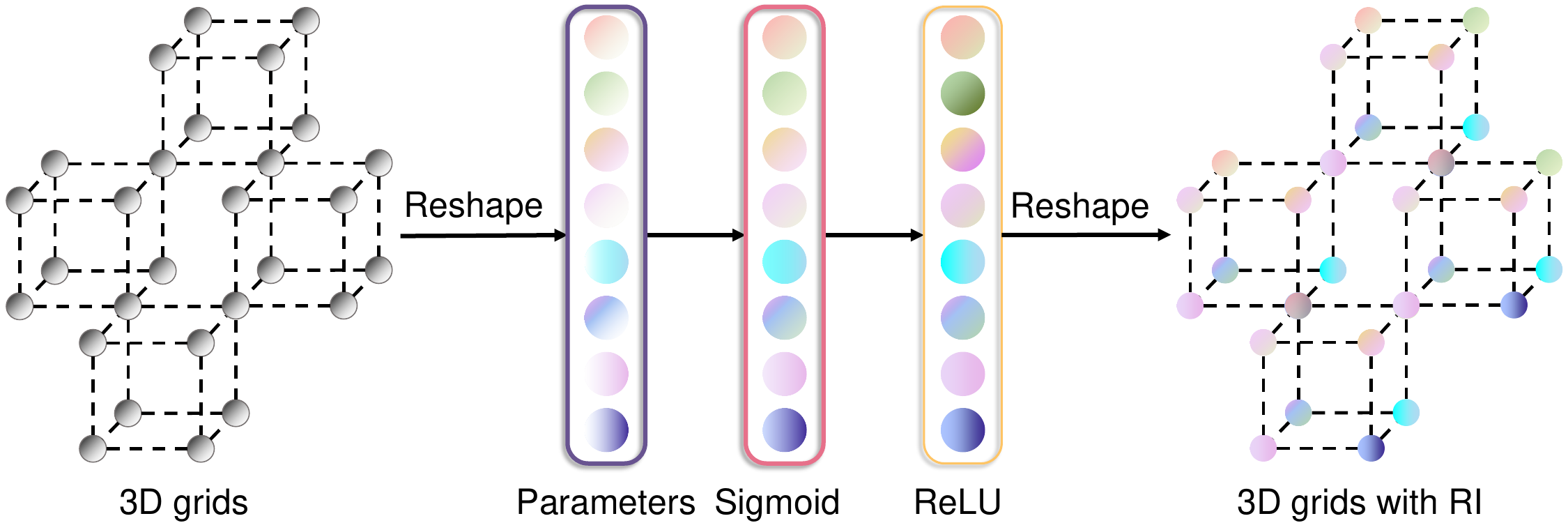}
\caption{\textbf{Structure of the explicit representation and neural field of FDT.} \textbf{a}, Initial 3D grids depicting interconnected nodes arranged in a spatial structure. \textbf{b}, Reshaping the 3D grids and setting them as the parameters of the neural field. \textbf{c}, Processing through parameter layers followed by sigmoid and ReLU functions. Mapping input values to a range between 0 and 1. \textbf{d}, Further reshaping the RI for the multi-slice model.}
\label{network}
\end{figure}

State-of-the-art methods use implicit neural networks \cite{liu2022recovery} to reconstruct the RI from intensity images, leveraging their ability to represent high-dimensional data. In our approach, the multi-slice model imposes a strong constraint on the RI. The multi-slice model enforces consistency across multiple 2D slices of the sample and ensures that the RI values are accurately aligned and correlated throughout the 3D volume. This inherent constraint reduces the complexity and ambiguity typically associated with high-dimensional data. Thus, instead of using implicit representations, we use explicit representations \cite{fridovich2022plenoxels} to model the 3D RI (Fig. \ref{overview}). Here, we provide more details about the framework (Fig. \ref{network}).

We first define a sparse voxel grid corresponding to the region of interest in the biological sample. Each voxel is then initialized as a network parameter using the Xavier method, representing the initial estimation of the unknown 3D RI. We then combine the explicit neural field with a coarse-to-fine structure, as described in Section \ref{subsec2.1}, to adaptively upsample the grid as the number of iterations increases. 

The explicit approach omits traditional neural network layers, such as linear or convolutional layers, which significantly accelerates the model's training speed. To prevent large or negative values, we apply a combination of sigmoid and ReLU activation functions to the parameters:

\begin{equation}
\text{Out} = \text{ReLU}(\text{Sigmoid}(x) - 0.5)
\end{equation}

After this transformation, the parameters are reshaped back into grid format and then fed into the differential multi-slice model.


\newpage

\section*{Data availability}
The data used for reproducing the results in the manuscript are available at \href{https://xue-lab-cobi.github.io/fdt/}{FDT website}
\section*{Code availability}
The code used for reproducing the results in the manuscript is available at \href{https://xue-lab-cobi.github.io/fdt/}{FDT website}
\section*{Acknowledgement}
Research reported in this publication was supported by the National Institute Of General Medical Sciences of the National Institutes of Health under Award Number R35GM155193. The content is solely the responsibility of the authors and does not necessarily represent the official views of the National Institutes of Health. This work was also supported by Dr. Yi Xue’s startup funds from the Department of Biomedical Engineering at the University of California, Davis.
We acknowledge Dr. Soichiro Yamada for providing the MDCK cells and biological support. We also acknowledge Dr. Jiandi Wan for providing the 3D cultured bovine myotube and mentoring Junjie Chen. 

\section*{Contributions}
The project was conceived by Y.X., and R.H.. The code of the model was implemented by R.H.. The experiments were designed by R.H., Y.C., and Y.X.. The numerical results were collected by R.H.. The 3D cultured bovine myotube was cultured by J.C.. The data acquisition and preparation was conducted by Y.C. and R.H.. The manuscript was primarily drafted by R.H. and revised by Y.X., and reviewed by all authors. 

\section*{Competing interests}
The authors declare no competing interests.






\clearpage
\begin{appendices}
\setcounter{figure}{0} 
\renewcommand{\thefigure}{S\arabic{figure}}
\renewcommand{\thetable}{S\arabic{table}}

\section{Evaluation Metrics and Visual Analysis for Simulated Data}\label{secA1}
\hypertarget{appendix}{}
\begin{table}[htbp]
    \centering
    \begin{tabular}{cc|cccc}
        \hline
        \multicolumn{2}{c|}{Simulated data} & MSE & SSIM & LPIPS & PSNR \\
        \hline
        \multirow{2}{*}{w/o coarse-to-fine} & images & 3.2253e-06 & 0.9994 & 0.0004 & 54.9143  \\
        & RI & 1.8282e-06 & 1.000 & 0.0010 & 57.3798  \\
        \hline
        \multirow{2}{*}{w/ coarse-to-fine } & 
          images & 6.1125e-07 & 0.9999 & 0.0001 & 62.1378  \\
        & RI & 1.5730e-06 & 1.000 & 0.0007 & 58.0327  \\
        \hline

        \multirow{2}{*}{\makecell{w/o self-calibration \\on parameters}} &  images & 1.0876e-03 & 0.8317 & 0.0720 & 29.6355  \\
        & RI & 7.7316e-05 & 0.9977 & 0.0063 & 41.1173  \\
        \hline
        \multirow{2}{*}{\makecell{w/ self-calibration \\on parameters}} &  images  & 5.6229e-06 & 0.9990 & 0.0006 & 52.5004   \\
        & RI & 5.2851e-06 & 1.0000 & 0.0019 & 52.7695   \\
        \hline

        \multirow{2}{*}{\makecell{w/o self-calibration \\on positions}} & images &  1.1377e-04 & 0.9723 & 0.0065 & 39.4398  \\
        & RI & 2.7980e-05 & 0.9994 & 0.0027 & 45.5315  \\
        \hline
        \multirow{2}{*}{\makecell{w/ self-calibration \\on positions}} & images & 3.1797e-05 & 0.9933 & 0.0052 & 44.9761  \\
        & RI & 3.1984e-06 & 0.9999 & 0.0007 & 54.9507  \\
        \hline
        
    \end{tabular}
    \caption{Quantitative results from the ablation study on the coarse-to-fine structure and self-calbiration using the simulated data ``UCDavis".}
    
    \label{tab:simulated_data2}
\end{table}

\begin{figure}[htbp]
\centering
\includegraphics[width=0.75\textwidth, trim={0 780 0 0},clip]{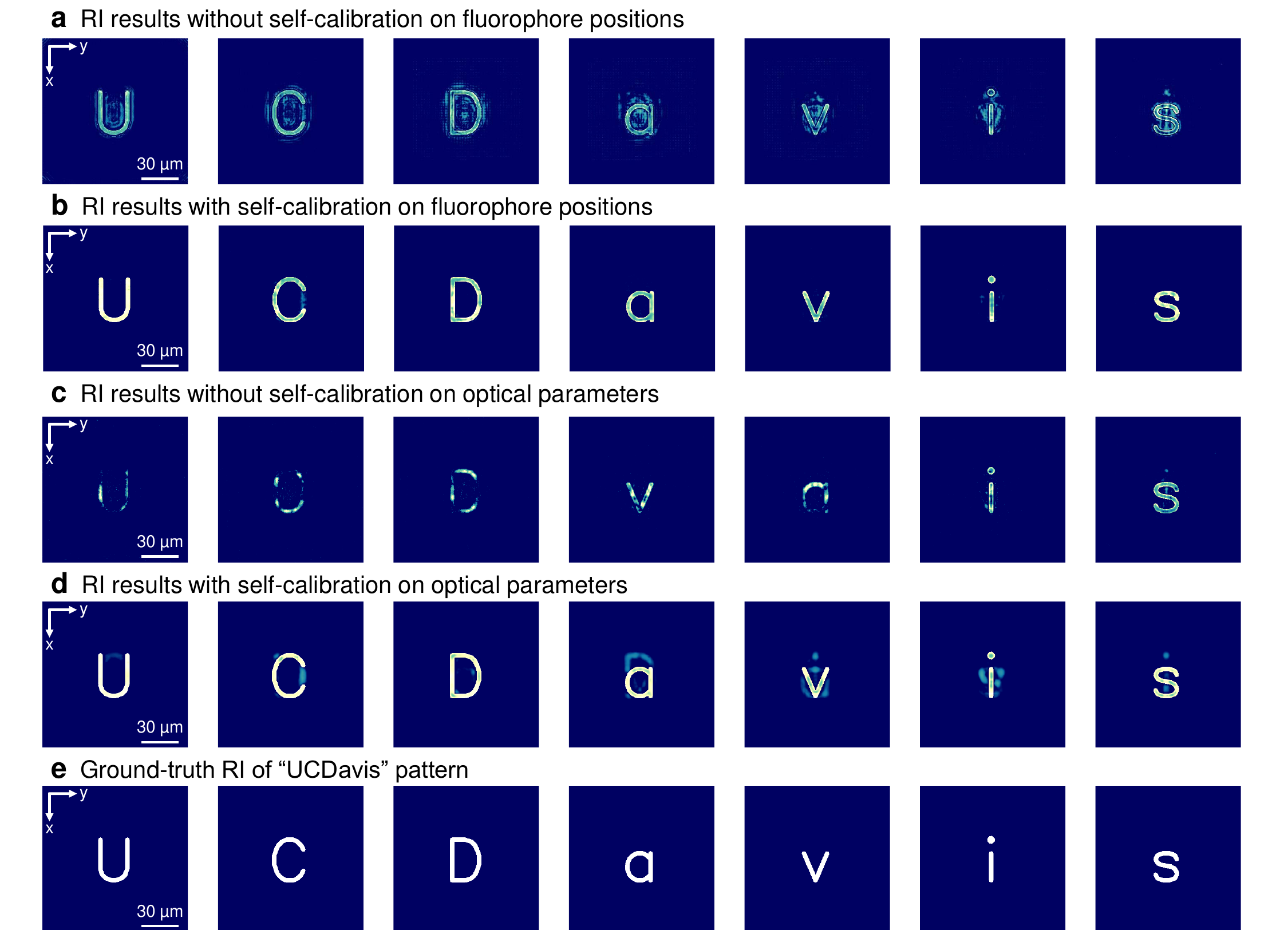}
\caption{\textbf{RI reconstruction of the ``UCDavis'' pattern with and without self-calibration.} \textbf{a}. Reconstructed RI without self-calibration on fluorophore positions shows significant distortion and noise. \textbf{b}. Reconstructed RI with self-calibration on fluorophore positions displays improved clarity and accuracy of the pattern. \textbf{c}. Reconstructed RI without self-calibration on optical parameters (e.g., the resolution of sampling grid) suffers from similar distortions seen in \textbf{a}. \textbf{d}. Reconstructed RI with self-calibration on optical parameters demonstrates further enhancement, closely matching the expected output. \textbf{e}. Ground-truth RI of the "UCDavis" pattern, showing the ideal pattern for comparison.}
\label{ucd_sup}
\end{figure}

\hypertarget{appendixfigS2}{}
\begin{figure}[htbp]
\centering
\includegraphics[width=1\textwidth, trim={0 580 0 0},clip]{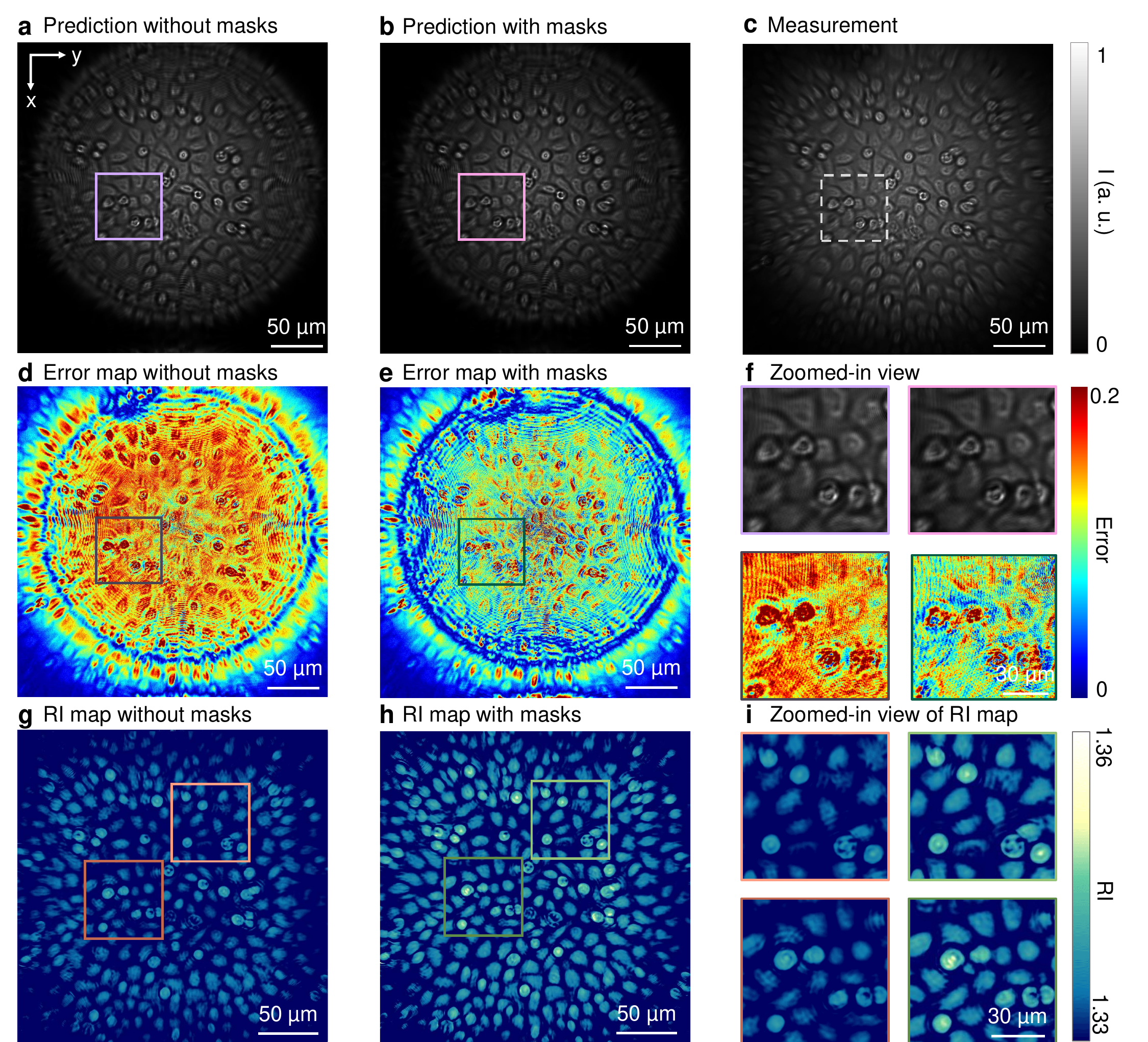}
\caption{\textbf{Comparison of reconstructed results using FDT trained with and without partially coherent masks.} 
\textbf{a-b}. Predicted image generated using the 3D RI calculated with FDT, trained \textbf{a} without and \textbf{b} with partially coherent masks.  
\textbf{c}. Experimentally measured image under the same illumination as in \textbf{a-b}. 
\textbf{d-e}. Error map between predicted image and measured images: \textbf{d} without and \textbf{e} with partially coherent masks. The partially coherent masks significantly reduce errors and improve accuracy. 
\textbf{f}. Zoomed-in view of representative regions in the images and error maps, confirming that the partially coherent masks improve the accuracy for both the bottom and top layer cells.   
\textbf{g-h}. A representative $z$-plane of the reconstructed RI: \textbf{g} without and \textbf{h} with the partially coherent masks, with 70 pixels cropped from each edge for better visualization.  
\textbf{i}. Zoomed-in view of representative regions in the RI maps.}
\label{mdck_sup}
\end{figure}

\clearpage


\end{appendices}


\bibliography{sn-bibliography}


\begin{thebibliography}{51}
\ifx \bisbn   \undefined \def \bisbn  #1{ISBN #1}\fi
\ifx \binits  \undefined \def \binits#1{#1}\fi
\ifx \bauthor  \undefined \def \bauthor#1{#1}\fi
\ifx \batitle  \undefined \def \batitle#1{#1}\fi
\ifx \bjtitle  \undefined \def \bjtitle#1{#1}\fi
\ifx \bvolume  \undefined \def \bvolume#1{\textbf{#1}}\fi
\ifx \byear  \undefined \def \byear#1{#1}\fi
\ifx \bissue  \undefined \def \bissue#1{#1}\fi
\ifx \bfpage  \undefined \def \bfpage#1{#1}\fi
\ifx \blpage  \undefined \def \blpage #1{#1}\fi
\ifx \burl  \undefined \def \burl#1{\textsf{#1}}\fi
\ifx \doiurl  \undefined \def \doiurl#1{\url{https://doi.org/#1}}\fi
\ifx \betal  \undefined \def \betal{\textit{et al.}}\fi
\ifx \binstitute  \undefined \def \binstitute#1{#1}\fi
\ifx \binstitutionaled  \undefined \def \binstitutionaled#1{#1}\fi
\ifx \bctitle  \undefined \def \bctitle#1{#1}\fi
\ifx \beditor  \undefined \def \beditor#1{#1}\fi
\ifx \bpublisher  \undefined \def \bpublisher#1{#1}\fi
\ifx \bbtitle  \undefined \def \bbtitle#1{#1}\fi
\ifx \bedition  \undefined \def \bedition#1{#1}\fi
\ifx \bseriesno  \undefined \def \bseriesno#1{#1}\fi
\ifx \blocation  \undefined \def \blocation#1{#1}\fi
\ifx \bsertitle  \undefined \def \bsertitle#1{#1}\fi
\ifx \bsnm \undefined \def \bsnm#1{#1}\fi
\ifx \bsuffix \undefined \def \bsuffix#1{#1}\fi
\ifx \bparticle \undefined \def \bparticle#1{#1}\fi
\ifx \barticle \undefined \def \barticle#1{#1}\fi
\bibcommenthead
\ifx \bconfdate \undefined \def \bconfdate #1{#1}\fi
\ifx \botherref \undefined \def \botherref #1{#1}\fi
\ifx \url \undefined \def \url#1{\textsf{#1}}\fi
\ifx \bchapter \undefined \def \bchapter#1{#1}\fi
\ifx \bbook \undefined \def \bbook#1{#1}\fi
\ifx \bcomment \undefined \def \bcomment#1{#1}\fi
\ifx \oauthor \undefined \def \oauthor#1{#1}\fi
\ifx \citeauthoryear \undefined \def \citeauthoryear#1{#1}\fi
\ifx \endbibitem  \undefined \def \endbibitem {}\fi
\ifx \bconflocation  \undefined \def \bconflocation#1{#1}\fi
\ifx \arxivurl  \undefined \def \arxivurl#1{\textsf{#1}}\fi
\csname PreBibitemsHook\endcsname

\bibitem[\protect\citeauthoryear{Xue et~al.}{2022}]{Xue:22}
\begin{barticle}
\bauthor{\bsnm{Xue}, \binits{Y.}},
\bauthor{\bsnm{Ren}, \binits{D.}},
\bauthor{\bsnm{Waller}, \binits{L.}}:
\batitle{Three-dimensional bi-functional refractive index and fluorescence microscopy (brief)}.
\bjtitle{Biomed. Opt. Express}
\bvolume{13}(\bissue{11}),
\bfpage{5900}--\blpage{5908}
(\byear{2022})
\doiurl{10.1364/BOE.456621}
\end{barticle}
\endbibitem

\bibitem[\protect\citeauthoryear{Li and Xue}{2024}]{2p-brief}
\begin{bchapter}
\bauthor{\bsnm{Li}, \binits{Y.}},
\bauthor{\bsnm{Xue}, \binits{Y.}}:
\bctitle{{Two-photon bi-functional refractive Index and fluorescence microscopy (2P-BRIEF)}}.
In: \beditor{\bsnm{Brown}, \binits{T.G.}},
\beditor{\bsnm{Wilson}, \binits{T.}},
\beditor{\bsnm{Waller}, \binits{L.}} (eds.)
\bbtitle{Three-Dimensional and Multidimensional Microscopy: Image Acquisition and Processing XXXI},
vol. \bseriesno{PC12848},
p. \bfpage{128480}.
\bpublisher{SPIE}, \blocation{???}
(\byear{2024}).
\doiurl{10.1117/12.3001958} .
\bcomment{International Society for Optics and Photonics}.
\burl{https://doi.org/10.1117/12.3001958}
\end{bchapter}
\endbibitem

\bibitem[\protect\citeauthoryear{Park et~al.}{2006}]{Park:06}
\begin{barticle}
\bauthor{\bsnm{Park}, \binits{Y.}},
\bauthor{\bsnm{Popescu}, \binits{G.}},
\bauthor{\bsnm{Badizadegan}, \binits{K.}},
\bauthor{\bsnm{Dasari}, \binits{R.R.}},
\bauthor{\bsnm{Feld}, \binits{M.S.}}:
\batitle{Diffraction phase and fluorescence microscopy}.
\bjtitle{Opt. Express}
\bvolume{14}(\bissue{18}),
\bfpage{8263}--\blpage{8268}
(\byear{2006})
\doiurl{10.1364/OE.14.008263}
\end{barticle}
\endbibitem

\bibitem[\protect\citeauthoryear{Kim et~al.}{2017}]{Kim:17}
\begin{barticle}
\bauthor{\bsnm{Kim}, \binits{K.}},
\bauthor{\bsnm{Park}, \binits{W.S.}},
\bauthor{\bsnm{Na}, \binits{S.}},
\bauthor{\bsnm{Kim}, \binits{S.}},
\bauthor{\bsnm{Kim}, \binits{T.}},
\bauthor{\bsnm{Heo}, \binits{W.D.}},
\bauthor{\bsnm{Park}, \binits{Y.}}:
\batitle{Correlative three-dimensional fluorescence and refractive index tomography: bridging the gap between molecular specificity and quantitative bioimaging}.
\bjtitle{Biomed. Opt. Express}
\bvolume{8}(\bissue{12}),
\bfpage{5688}--\blpage{5697}
(\byear{2017})
\doiurl{10.1364/BOE.8.005688}
\end{barticle}
\endbibitem

\bibitem[\protect\citeauthoryear{Chowdhury et~al.}{2017}]{Chowdhury:17}
\begin{barticle}
\bauthor{\bsnm{Chowdhury}, \binits{S.}},
\bauthor{\bsnm{Eldridge}, \binits{W.J.}},
\bauthor{\bsnm{Wax}, \binits{A.}},
\bauthor{\bsnm{Izatt}, \binits{J.A.}}:
\batitle{Structured illumination multimodal 3d-resolved quantitative phase and fluorescence sub-diffraction microscopy}.
\bjtitle{Biomed. Opt. Express}
\bvolume{8}(\bissue{5}),
\bfpage{2496}--\blpage{2518}
(\byear{2017})
\doiurl{10.1364/BOE.8.002496}
\end{barticle}
\endbibitem

\bibitem[\protect\citeauthoryear{Yeh et~al.}{2019}]{Yeh:19}
\begin{barticle}
\bauthor{\bsnm{Yeh}, \binits{L.-H.}},
\bauthor{\bsnm{Chowdhury}, \binits{S.}},
\bauthor{\bsnm{Waller}, \binits{L.}}:
\batitle{Computational structured illumination for high-content fluorescence and phase microscopy}.
\bjtitle{Biomed. Opt. Express}
\bvolume{10}(\bissue{4}),
\bfpage{1978}--\blpage{1998}
(\byear{2019})
\doiurl{10.1364/BOE.10.001978}
\end{barticle}
\endbibitem

\bibitem[\protect\citeauthoryear{Dong et~al.}{2020}]{Dong2020-dq}
\begin{barticle}
\bauthor{\bsnm{Dong}, \binits{D.}},
\bauthor{\bsnm{Huang}, \binits{X.}},
\bauthor{\bsnm{Li}, \binits{L.}},
\bauthor{\bsnm{Mao}, \binits{H.}},
\bauthor{\bsnm{Mo}, \binits{Y.}},
\bauthor{\bsnm{Zhang}, \binits{G.}},
\bauthor{\bsnm{Zhang}, \binits{Z.}},
\bauthor{\bsnm{Shen}, \binits{J.}},
\bauthor{\bsnm{Liu}, \binits{W.}},
\bauthor{\bsnm{Wu}, \binits{Z.}},
\bauthor{\bsnm{Liu}, \binits{G.}},
\bauthor{\bsnm{Liu}, \binits{Y.}},
\bauthor{\bsnm{Yang}, \binits{H.}},
\bauthor{\bsnm{Gong}, \binits{Q.}},
\bauthor{\bsnm{Shi}, \binits{K.}},
\bauthor{\bsnm{Chen}, \binits{L.}}:
\batitle{Super-resolution fluorescence-assisted diffraction computational tomography reveals the three-dimensional landscape of the cellular organelle interactome}.
\bjtitle{Light Sci Appl}
\bvolume{9},
\bfpage{11}
(\byear{2020})
\end{barticle}
\endbibitem

\bibitem[\protect\citeauthoryear{Shaffer et~al.}{2012}]{shaffer2012single}
\begin{barticle}
\bauthor{\bsnm{Shaffer}, \binits{E.}},
\bauthor{\bsnm{Pavillon}, \binits{N.}},
\bauthor{\bsnm{Depeursinge}, \binits{C.}}:
\batitle{Single-shot, simultaneous incoherent and holographic microscopy}.
\bjtitle{Journal of microscopy}
\bvolume{245}(\bissue{1}),
\bfpage{49}--\blpage{62}
(\byear{2012})
\end{barticle}
\endbibitem

\bibitem[\protect\citeauthoryear{Marthy et~al.}{2024}]{marthy2024single}
\begin{barticle}
\bauthor{\bsnm{Marthy}, \binits{B.}},
\bauthor{\bsnm{B{\'e}n{\'e}fice}, \binits{M.}},
\bauthor{\bsnm{Baffou}, \binits{G.}}:
\batitle{Single-shot quantitative phase-fluorescence imaging using cross-grating wavefront microscopy}.
\bjtitle{Scientific Reports}
\bvolume{14}(\bissue{1}),
\bfpage{2142}
(\byear{2024})
\end{barticle}
\endbibitem

\bibitem[\protect\citeauthoryear{Quan et~al.}{2021}]{9260959}
\begin{barticle}
\bauthor{\bsnm{Quan}, \binits{X.}},
\bauthor{\bsnm{Kumar}, \binits{M.}},
\bauthor{\bsnm{Rajput}, \binits{S.K.}},
\bauthor{\bsnm{Tamada}, \binits{Y.}},
\bauthor{\bsnm{Awatsuji}, \binits{Y.}},
\bauthor{\bsnm{Matoba}, \binits{O.}}:
\batitle{Multimodal microscopy: Fast acquisition of quantitative phase and fluorescence imaging in 3d space}.
\bjtitle{IEEE Journal of Selected Topics in Quantum Electronics}
\bvolume{27}(\bissue{4}),
\bfpage{1}--\blpage{11}
(\byear{2021})
\doiurl{10.1109/JSTQE.2020.3038403}
\end{barticle}
\endbibitem

\bibitem[\protect\citeauthoryear{Tayal et~al.}{2020}]{tayal2020simultaneous}
\begin{barticle}
\bauthor{\bsnm{Tayal}, \binits{S.}},
\bauthor{\bsnm{Singh}, \binits{V.}},
\bauthor{\bsnm{Kaur}, \binits{T.}},
\bauthor{\bsnm{Singh}, \binits{N.}},
\bauthor{\bsnm{Mehta}, \binits{D.S.}}:
\batitle{Simultaneous fluorescence and quantitative phase imaging of mg63 osteosarcoma cells to monitor morphological changes with time using partially spatially coherent light source}.
\bjtitle{Methods and Applications in Fluorescence}
\bvolume{8}(\bissue{3}),
\bfpage{035004}
(\byear{2020})
\end{barticle}
\endbibitem

\bibitem[\protect\citeauthoryear{Rajput et~al.}{2021}]{9376596}
\begin{barticle}
\bauthor{\bsnm{Rajput}, \binits{S.K.}},
\bauthor{\bsnm{Matoba}, \binits{O.}},
\bauthor{\bsnm{Kumar}, \binits{M.}},
\bauthor{\bsnm{Quan}, \binits{X.}},
\bauthor{\bsnm{Awatsuji}, \binits{Y.}},
\bauthor{\bsnm{Tamada}, \binits{Y.}},
\bauthor{\bsnm{Tajahuerce}, \binits{E.}}:
\batitle{Multi-physical parameter cross-sectional imaging of quantitative phase and fluorescence by integrated multimodal microscopy}.
\bjtitle{IEEE Journal of Selected Topics in Quantum Electronics}
\bvolume{27}(\bissue{4}),
\bfpage{1}--\blpage{9}
(\byear{2021})
\doiurl{10.1109/JSTQE.2021.3064406}
\end{barticle}
\endbibitem

\bibitem[\protect\citeauthoryear{Liu et~al.}{2018}]{Liu:18}
\begin{barticle}
\bauthor{\bsnm{Liu}, \binits{Y.}},
\bauthor{\bsnm{Suo}, \binits{J.}},
\bauthor{\bsnm{Zhang}, \binits{Y.}},
\bauthor{\bsnm{Dai}, \binits{Q.}}:
\batitle{Single-pixel phase and fluorescence microscope}.
\bjtitle{Opt. Express}
\bvolume{26}(\bissue{25}),
\bfpage{32451}--\blpage{32462}
(\byear{2018})
\doiurl{10.1364/OE.26.032451}
\end{barticle}
\endbibitem

\bibitem[\protect\citeauthoryear{Pavillon et~al.}{2010}]{pavillon2010cell}
\begin{barticle}
\bauthor{\bsnm{Pavillon}, \binits{N.}},
\bauthor{\bsnm{Benke}, \binits{A.}},
\bauthor{\bsnm{Boss}, \binits{D.}},
\bauthor{\bsnm{Moratal}, \binits{C.}},
\bauthor{\bsnm{K{\"u}hn}, \binits{J.}},
\bauthor{\bsnm{Jourdain}, \binits{P.}},
\bauthor{\bsnm{Depeursinge}, \binits{C.}},
\bauthor{\bsnm{Magistretti}, \binits{P.J.}},
\bauthor{\bsnm{Marquet}, \binits{P.}}:
\batitle{Cell morphology and intracellular ionic homeostasis explored with a multimodal approach combining epifluorescence and digital holographic microscopy}.
\bjtitle{Journal of biophotonics}
\bvolume{3}(\bissue{7}),
\bfpage{432}--\blpage{436}
(\byear{2010})
\end{barticle}
\endbibitem

\bibitem[\protect\citeauthoryear{an~Pham et~al.}{2021}]{PHAM2021127290}
\begin{barticle}
\bauthor{\bsnm{Pham}, \binits{T.-a.}},
\bauthor{\bsnm{Soubies}, \binits{E.}},
\bauthor{\bsnm{Soulez}, \binits{F.}},
\bauthor{\bsnm{Unser}, \binits{M.}}:
\batitle{Optical diffraction tomography from single-molecule localization microscopy}.
\bjtitle{Optics Communications}
\bvolume{499},
\bfpage{127290}
(\byear{2021})
\doiurl{10.1016/j.optcom.2021.127290}
\end{barticle}
\endbibitem

\bibitem[\protect\citeauthoryear{Choi et~al.}{2007}]{Choi2007-ay}
\begin{barticle}
\bauthor{\bsnm{Choi}, \binits{W.}},
\bauthor{\bsnm{Fang-Yen}, \binits{C.}},
\bauthor{\bsnm{Badizadegan}, \binits{K.}},
\bauthor{\bsnm{Oh}, \binits{S.}},
\bauthor{\bsnm{Lue}, \binits{N.}},
\bauthor{\bsnm{Dasari}, \binits{R.R.}},
\bauthor{\bsnm{Feld}, \binits{M.S.}}:
\batitle{Tomographic phase microscopy}.
\bjtitle{Nat. Methods}
\bvolume{4}(\bissue{9}),
\bfpage{717}--\blpage{719}
(\byear{2007})
\end{barticle}
\endbibitem

\bibitem[\protect\citeauthoryear{Sung et~al.}{2009}]{Sung:09}
\begin{barticle}
\bauthor{\bsnm{Sung}, \binits{Y.}},
\bauthor{\bsnm{Choi}, \binits{W.}},
\bauthor{\bsnm{Fang-Yen}, \binits{C.}},
\bauthor{\bsnm{Badizadegan}, \binits{K.}},
\bauthor{\bsnm{Dasari}, \binits{R.R.}},
\bauthor{\bsnm{Feld}, \binits{M.S.}}:
\batitle{Optical diffraction tomography for high resolution live cell imaging}.
\bjtitle{Opt. Express}
\bvolume{17}(\bissue{1}),
\bfpage{266}--\blpage{277}
(\byear{2009})
\doiurl{10.1364/OE.17.000266}
\end{barticle}
\endbibitem

\bibitem[\protect\citeauthoryear{Waller et~al.}{2010}]{Waller2010-oj}
\begin{barticle}
\bauthor{\bsnm{Waller}, \binits{L.}},
\bauthor{\bsnm{Tian}, \binits{L.}},
\bauthor{\bsnm{Barbastathis}, \binits{G.}}:
\batitle{Transport of intensity phase-amplitude imaging with higher order intensity derivatives}.
\bjtitle{Opt. Express}
\bvolume{18}(\bissue{12}),
\bfpage{12552}--\blpage{12561}
(\byear{2010})
\end{barticle}
\endbibitem

\bibitem[\protect\citeauthoryear{Tian and Waller}{2015}]{Tian2015-ij}
\begin{barticle}
\bauthor{\bsnm{Tian}, \binits{L.}},
\bauthor{\bsnm{Waller}, \binits{L.}}:
\batitle{{3D} intensity and phase imaging from light field measurements in an {LED} array microscope}.
\bjtitle{Optica, OPTICA}
\bvolume{2}(\bissue{2}),
\bfpage{104}--\blpage{111}
(\byear{2015})
\end{barticle}
\endbibitem

\bibitem[\protect\citeauthoryear{Choi et~al.}{2014}]{Choi2014-rd}
\begin{barticle}
\bauthor{\bsnm{Choi}, \binits{Y.}},
\bauthor{\bsnm{Hosseini}, \binits{P.}},
\bauthor{\bsnm{Choi}, \binits{W.}},
\bauthor{\bsnm{Dasari}, \binits{R.R.}},
\bauthor{\bsnm{So}, \binits{P.T.C.}},
\bauthor{\bsnm{Yaqoob}, \binits{Z.}}:
\batitle{Dynamic speckle illumination wide-field reflection phase microscopy}.
\bjtitle{Opt. Lett.}
\bvolume{39}(\bissue{20}),
\bfpage{6062}--\blpage{6065}
(\byear{2014})
\end{barticle}
\endbibitem

\bibitem[\protect\citeauthoryear{Kang et~al.}{2015}]{Kang2015-mz}
\begin{barticle}
\bauthor{\bsnm{Kang}, \binits{S.}},
\bauthor{\bsnm{Jeong}, \binits{S.}},
\bauthor{\bsnm{Choi}, \binits{W.}},
\bauthor{\bsnm{Ko}, \binits{H.}},
\bauthor{\bsnm{Yang}, \binits{T.D.}},
\bauthor{\bsnm{Joo}, \binits{J.H.}},
\bauthor{\bsnm{Lee}, \binits{J.-S.}},
\bauthor{\bsnm{Lim}, \binits{Y.-S.}},
\bauthor{\bsnm{Park}, \binits{Q.-H.}},
\bauthor{\bsnm{Choi}, \binits{W.}}:
\batitle{Imaging deep within a scattering medium using collective accumulation of single-scattered waves}.
\bjtitle{Nat. Photonics}
\bvolume{9},
\bfpage{253}
(\byear{2015})
\end{barticle}
\endbibitem

\bibitem[\protect\citeauthoryear{Singh et~al.}{2019}]{Singh2019-zy}
\begin{barticle}
\bauthor{\bsnm{Singh}, \binits{V.R.}},
\bauthor{\bsnm{Yang}, \binits{Y.A.}},
\bauthor{\bsnm{Yu}, \binits{H.}},
\bauthor{\bsnm{Kamm}, \binits{R.D.}},
\bauthor{\bsnm{Yaqoob}, \binits{Z.}},
\bauthor{\bsnm{So}, \binits{P.T.C.}}:
\batitle{Studying nucleic envelope and plasma membrane mechanics of eukaryotic cells using confocal reflectance interferometric microscopy}.
\bjtitle{Nat. Commun.}
\bvolume{10}(\bissue{1}),
\bfpage{3652}
(\byear{2019})
\end{barticle}
\endbibitem

\bibitem[\protect\citeauthoryear{Hyeon et~al.}{2021}]{hyeon2021effect}
\begin{barticle}
\bauthor{\bsnm{Hyeon}, \binits{M.G.}},
\bauthor{\bsnm{Park}, \binits{K.}},
\bauthor{\bsnm{Yang}, \binits{T.D.}},
\bauthor{\bsnm{Kong}, \binits{T.}},
\bauthor{\bsnm{Kim}, \binits{B.-M.}},
\bauthor{\bsnm{Choi}, \binits{Y.}}:
\batitle{The effect of pupil transmittance on axial resolution of reflection phase microscopy}.
\bjtitle{Scientific reports}
\bvolume{11}(\bissue{1}),
\bfpage{22774}
(\byear{2021})
\end{barticle}
\endbibitem

\bibitem[\protect\citeauthoryear{Kang et~al.}{2023}]{Kang:23}
\begin{barticle}
\bauthor{\bsnm{Kang}, \binits{Y.G.}},
\bauthor{\bsnm{Park}, \binits{K.}},
\bauthor{\bsnm{Hyeon}, \binits{M.G.}},
\bauthor{\bsnm{Yang}, \binits{T.D.}},
\bauthor{\bsnm{Choi}, \binits{Y.}}:
\batitle{Three-dimensional imaging in reflection phase microscopy with minimal axial scanning}.
\bjtitle{Opt. Express}
\bvolume{31}(\bissue{26}),
\bfpage{44741}--\blpage{44753}
(\byear{2023})
\doiurl{10.1364/OE.510519}
\end{barticle}
\endbibitem

\bibitem[\protect\citeauthoryear{Wang et~al.}{2024}]{wang2024use}
\begin{barticle}
\bauthor{\bsnm{Wang}, \binits{K.}},
\bauthor{\bsnm{Song}, \binits{L.}},
\bauthor{\bsnm{Wang}, \binits{C.}},
\bauthor{\bsnm{Ren}, \binits{Z.}},
\bauthor{\bsnm{Zhao}, \binits{G.}},
\bauthor{\bsnm{Dou}, \binits{J.}},
\bauthor{\bsnm{Di}, \binits{J.}},
\bauthor{\bsnm{Barbastathis}, \binits{G.}},
\bauthor{\bsnm{Zhou}, \binits{R.}},
\bauthor{\bsnm{Zhao}, \binits{J.}}, \betal:
\batitle{On the use of deep learning for phase recovery}.
\bjtitle{Light: Science \& Applications}
\bvolume{13}(\bissue{1}),
\bfpage{4}
(\byear{2024})
\end{barticle}
\endbibitem

\bibitem[\protect\citeauthoryear{Dong et~al.}{2023}]{10004797}
\begin{barticle}
\bauthor{\bsnm{Dong}, \binits{J.}},
\bauthor{\bsnm{Valzania}, \binits{L.}},
\bauthor{\bsnm{Maillard}, \binits{A.}},
\bauthor{\bsnm{Pham}, \binits{T.-a.}},
\bauthor{\bsnm{Gigan}, \binits{S.}},
\bauthor{\bsnm{Unser}, \binits{M.}}:
\batitle{Phase retrieval: From computational imaging to machine learning: A tutorial}.
\bjtitle{IEEE Signal Processing Magazine}
\bvolume{40}(\bissue{1}),
\bfpage{45}--\blpage{57}
(\byear{2023})
\doiurl{10.1109/MSP.2022.3219240}
\end{barticle}
\endbibitem

\bibitem[\protect\citeauthoryear{Kamilov et~al.}{2015}]{Kamilov:15}
\begin{barticle}
\bauthor{\bsnm{Kamilov}, \binits{U.S.}},
\bauthor{\bsnm{Papadopoulos}, \binits{I.N.}},
\bauthor{\bsnm{Shoreh}, \binits{M.H.}},
\bauthor{\bsnm{Goy}, \binits{A.}},
\bauthor{\bsnm{Vonesch}, \binits{C.}},
\bauthor{\bsnm{Unser}, \binits{M.}},
\bauthor{\bsnm{Psaltis}, \binits{D.}}:
\batitle{Learning approach to optical tomography}.
\bjtitle{Optica}
\bvolume{2}(\bissue{6}),
\bfpage{517}--\blpage{522}
(\byear{2015})
\doiurl{10.1364/OPTICA.2.000517}
\end{barticle}
\endbibitem

\bibitem[\protect\citeauthoryear{Wu et~al.}{2022}]{Wu:22}
\begin{barticle}
\bauthor{\bsnm{Wu}, \binits{X.}},
\bauthor{\bsnm{Wu}, \binits{Z.}},
\bauthor{\bsnm{Shanmugavel}, \binits{S.C.}},
\bauthor{\bsnm{Yu}, \binits{H.Z.}},
\bauthor{\bsnm{Zhu}, \binits{Y.}}:
\batitle{Physics-informed neural network for phase imaging based on transport of intensity equation}.
\bjtitle{Opt. Express}
\bvolume{30}(\bissue{24}),
\bfpage{43398}--\blpage{43416}
(\byear{2022})
\doiurl{10.1364/OE.462844}
\end{barticle}
\endbibitem

\bibitem[\protect\citeauthoryear{Matlock et~al.}{2023}]{Matlock:23}
\begin{barticle}
\bauthor{\bsnm{Matlock}, \binits{A.}},
\bauthor{\bsnm{Zhu}, \binits{J.}},
\bauthor{\bsnm{Tian}, \binits{L.}}:
\batitle{Multiple-scattering simulator-trained neural network for intensity diffraction tomography}.
\bjtitle{Opt. Express}
\bvolume{31}(\bissue{3}),
\bfpage{4094}--\blpage{4107}
(\byear{2023})
\doiurl{10.1364/OE.477396}
\end{barticle}
\endbibitem

\bibitem[\protect\citeauthoryear{Zhou and Horstmeyer}{2020}]{Zhou:20}
\begin{barticle}
\bauthor{\bsnm{Zhou}, \binits{K.C.}},
\bauthor{\bsnm{Horstmeyer}, \binits{R.}}:
\batitle{Diffraction tomography with a deep image prior}.
\bjtitle{Opt. Express}
\bvolume{28}(\bissue{9}),
\bfpage{12872}--\blpage{12896}
(\byear{2020})
\doiurl{10.1364/OE.379200}
\end{barticle}
\endbibitem

\bibitem[\protect\citeauthoryear{Raissi et~al.}{2019}]{Raissi2019-el}
\begin{barticle}
\bauthor{\bsnm{Raissi}, \binits{M.}},
\bauthor{\bsnm{Perdikaris}, \binits{P.}},
\bauthor{\bsnm{Karniadakis}, \binits{G.E.}}:
\batitle{Physics-informed neural networks: A deep learning framework for solving forward and inverse problems involving nonlinear partial differential equations}.
\bjtitle{J. Comput. Phys.}
\bvolume{378},
\bfpage{686}--\blpage{707}
(\byear{2019})
\end{barticle}
\endbibitem

\bibitem[\protect\citeauthoryear{Saba et~al.}{2022}]{saba2022physics}
\begin{barticle}
\bauthor{\bsnm{Saba}, \binits{A.}},
\bauthor{\bsnm{Gigli}, \binits{C.}},
\bauthor{\bsnm{Ayoub}, \binits{A.B.}},
\bauthor{\bsnm{Psaltis}, \binits{D.}}:
\batitle{Physics-informed neural networks for diffraction tomography}.
\bjtitle{Advanced Photonics}
\bvolume{4}(\bissue{6}),
\bfpage{066001}--\blpage{066001}
(\byear{2022})
\end{barticle}
\endbibitem

\bibitem[\protect\citeauthoryear{Yang et~al.}{2023}]{Yang:23}
\begin{barticle}
\bauthor{\bsnm{Yang}, \binits{D.}},
\bauthor{\bsnm{Zhang}, \binits{S.}},
\bauthor{\bsnm{Zheng}, \binits{C.}},
\bauthor{\bsnm{Zhou}, \binits{G.}},
\bauthor{\bsnm{Hu}, \binits{Y.}},
\bauthor{\bsnm{Hao}, \binits{Q.}}:
\batitle{Refractive index tomography with a physics-based optical neural network}.
\bjtitle{Biomed. Opt. Express}
\bvolume{14}(\bissue{11}),
\bfpage{5886}--\blpage{5903}
(\byear{2023})
\doiurl{10.1364/BOE.504242}
\end{barticle}
\endbibitem

\bibitem[\protect\citeauthoryear{Xu et~al.}{2022}]{xu2022point}
\begin{bchapter}
\bauthor{\bsnm{Xu}, \binits{Q.}},
\bauthor{\bsnm{Xu}, \binits{Z.}},
\bauthor{\bsnm{Philip}, \binits{J.}},
\bauthor{\bsnm{Bi}, \binits{S.}},
\bauthor{\bsnm{Shu}, \binits{Z.}},
\bauthor{\bsnm{Sunkavalli}, \binits{K.}},
\bauthor{\bsnm{Neumann}, \binits{U.}}:
\bctitle{Point-nerf: Point-based neural radiance fields}.
In: \bbtitle{Proceedings of the IEEE/CVF Conference on Computer Vision and Pattern Recognition},
pp. \bfpage{5438}--\blpage{5448}
(\byear{2022})
\end{bchapter}
\endbibitem

\bibitem[\protect\citeauthoryear{Rzepecki et~al.}{2022}]{rzepecki2022fast}
\begin{botherref}
\oauthor{\bsnm{Rzepecki}, \binits{J.}},
\oauthor{\bsnm{Bates}, \binits{D.}},
\oauthor{\bsnm{Doran}, \binits{C.}}:
Fast neural network based solving of partial differential equations.
arXiv preprint arXiv:2205.08978
(2022)
\end{botherref}
\endbibitem

\bibitem[\protect\citeauthoryear{Zhou et~al.}{2023}]{Zhou:23}
\begin{barticle}
\bauthor{\bsnm{Zhou}, \binits{H.}},
\bauthor{\bsnm{Feng}, \binits{B.Y.}},
\bauthor{\bsnm{Guo}, \binits{H.}},
\bauthor{\bsnm{Lin}, \binits{S.S.}},
\bauthor{\bsnm{Liang}, \binits{M.}},
\bauthor{\bsnm{Metzler}, \binits{C.A.}},
\bauthor{\bsnm{Yang}, \binits{C.}}:
\batitle{Fourier ptychographic microscopy image stack reconstruction using implicit neural representations}.
\bjtitle{Optica}
\bvolume{10}(\bissue{12}),
\bfpage{1679}--\blpage{1687}
(\byear{2023})
\doiurl{10.1364/OPTICA.505283}
\end{barticle}
\endbibitem

\bibitem[\protect\citeauthoryear{Zhang et~al.}{2024}]{zhang2024single}
\begin{botherref}
\oauthor{\bsnm{Zhang}, \binits{O.}},
\oauthor{\bsnm{Zhou}, \binits{H.}},
\oauthor{\bsnm{Feng}, \binits{B.Y.}},
\oauthor{\bsnm{Larsson}, \binits{E.M.}},
\oauthor{\bsnm{Alcalde}, \binits{R.E.}},
\oauthor{\bsnm{Yin}, \binits{S.}},
\oauthor{\bsnm{Deng}, \binits{C.}},
\oauthor{\bsnm{Yang}, \binits{C.}}:
Single-shot volumetric fluorescence imaging with neural fields.
arXiv preprint arXiv:2405.10463
(2024)
\end{botherref}
\endbibitem

\bibitem[\protect\citeauthoryear{Kang et~al.}{2024}]{kang2024coordinate}
\begin{botherref}
\oauthor{\bsnm{Kang}, \binits{I.}},
\oauthor{\bsnm{Zhang}, \binits{Q.}},
\oauthor{\bsnm{Yu}, \binits{S.X.}},
\oauthor{\bsnm{Ji}, \binits{N.}}:
Coordinate-based neural representations for computational adaptive optics in widefield microscopy.
Nature Machine Intelligence,
1--12
(2024)
\end{botherref}
\endbibitem

\bibitem[\protect\citeauthoryear{Feng et~al.}{2023}]{Feng2023-ln}
\begin{barticle}
\bauthor{\bsnm{Feng}, \binits{B.Y.}},
\bauthor{\bsnm{Guo}, \binits{H.}},
\bauthor{\bsnm{Xie}, \binits{M.}},
\bauthor{\bsnm{Boominathan}, \binits{V.}},
\bauthor{\bsnm{Sharma}, \binits{M.K.}},
\bauthor{\bsnm{Veeraraghavan}, \binits{A.}},
\bauthor{\bsnm{Metzler}, \binits{C.A.}}:
\batitle{{NeuWS}: Neural wavefront shaping for guidestar-free imaging through static and dynamic scattering media}.
\bjtitle{Sci. Adv.}
\bvolume{9}(\bissue{26}),
\bfpage{4671}
(\byear{2023})
\end{barticle}
\endbibitem

\bibitem[\protect\citeauthoryear{Sun et~al.}{2021}]{sun2021coil}
\begin{barticle}
\bauthor{\bsnm{Sun}, \binits{Y.}},
\bauthor{\bsnm{Liu}, \binits{J.}},
\bauthor{\bsnm{Xie}, \binits{M.}},
\bauthor{\bsnm{Wohlberg}, \binits{B.}},
\bauthor{\bsnm{Kamilov}, \binits{U.S.}}:
\batitle{Coil: Coordinate-based internal learning for tomographic imaging}.
\bjtitle{IEEE Transactions on Computational Imaging}
\bvolume{7},
\bfpage{1400}--\blpage{1412}
(\byear{2021})
\end{barticle}
\endbibitem

\bibitem[\protect\citeauthoryear{Cao et~al.}{2022}]{cao2022dynamic}
\begin{bchapter}
\bauthor{\bsnm{Cao}, \binits{R.}},
\bauthor{\bsnm{Liu}, \binits{F.L.}},
\bauthor{\bsnm{Yeh}, \binits{L.-H.}},
\bauthor{\bsnm{Waller}, \binits{L.}}:
\bctitle{Dynamic structured illumination microscopy with a neural space-time model}.
In: \bbtitle{2022 IEEE International Conference on Computational Photography (ICCP)},
pp. \bfpage{1}--\blpage{12}
(\byear{2022}).
\bcomment{IEEE}
\end{bchapter}
\endbibitem

\bibitem[\protect\citeauthoryear{Liu et~al.}{2022}]{liu2022recovery}
\begin{barticle}
\bauthor{\bsnm{Liu}, \binits{R.}},
\bauthor{\bsnm{Sun}, \binits{Y.}},
\bauthor{\bsnm{Zhu}, \binits{J.}},
\bauthor{\bsnm{Tian}, \binits{L.}},
\bauthor{\bsnm{Kamilov}, \binits{U.S.}}:
\batitle{Recovery of continuous 3d refractive index maps from discrete intensity-only measurements using neural fields}.
\bjtitle{Nature Machine Intelligence}
\bvolume{4}(\bissue{9}),
\bfpage{781}--\blpage{791}
(\byear{2022})
\end{barticle}
\endbibitem

\bibitem[\protect\citeauthoryear{Fridovich-Keil et~al.}{2022}]{fridovich2022plenoxels}
\begin{bchapter}
\bauthor{\bsnm{Fridovich-Keil}, \binits{S.}},
\bauthor{\bsnm{Yu}, \binits{A.}},
\bauthor{\bsnm{Tancik}, \binits{M.}},
\bauthor{\bsnm{Chen}, \binits{Q.}},
\bauthor{\bsnm{Recht}, \binits{B.}},
\bauthor{\bsnm{Kanazawa}, \binits{A.}}:
\bctitle{Plenoxels: Radiance fields without neural networks}.
In: \bbtitle{Proceedings of the IEEE/CVF Conference on Computer Vision and Pattern Recognition},
pp. \bfpage{5501}--\blpage{5510}
(\byear{2022})
\end{bchapter}
\endbibitem

\bibitem[\protect\citeauthoryear{Zhang et~al.}{2020}]{zhang2020nerf++}
\begin{botherref}
\oauthor{\bsnm{Zhang}, \binits{K.}},
\oauthor{\bsnm{Riegler}, \binits{G.}},
\oauthor{\bsnm{Snavely}, \binits{N.}},
\oauthor{\bsnm{Koltun}, \binits{V.}}:
Nerf++: Analyzing and improving neural radiance fields.
arXiv preprint arXiv:2010.07492
(2020)
\end{botherref}
\endbibitem

\bibitem[\protect\citeauthoryear{Xue et~al.}{2019}]{Xue2019-fk}
\begin{barticle}
\bauthor{\bsnm{Xue}, \binits{Y.}},
\bauthor{\bsnm{Berry}, \binits{K.P.}},
\bauthor{\bsnm{Boivin}, \binits{J.R.}},
\bauthor{\bsnm{Rowlands}, \binits{C.J.}},
\bauthor{\bsnm{Takiguchi}, \binits{Y.}},
\bauthor{\bsnm{Nedivi}, \binits{E.}},
\bauthor{\bsnm{So}, \binits{P.T.C.}}:
\batitle{Scanless volumetric imaging by selective access multifocal multiphoton microscopy}.
\bjtitle{Optica}
\bvolume{6}(\bissue{1}),
\bfpage{76}--\blpage{83}
(\byear{2019})
\end{barticle}
\endbibitem

\bibitem[\protect\citeauthoryear{Bastiaans}{1986}]{bastiaans1986application}
\begin{barticle}
\bauthor{\bsnm{Bastiaans}, \binits{M.J.}}:
\batitle{Application of the wigner distribution function to partially coherent light}.
\bjtitle{JOSA A}
\bvolume{3}(\bissue{8}),
\bfpage{1227}--\blpage{1238}
(\byear{1986})
\end{barticle}
\endbibitem

\bibitem[\protect\citeauthoryear{Zuo et~al.}{2015}]{Zuo2015-or}
\begin{barticle}
\bauthor{\bsnm{Zuo}, \binits{C.}},
\bauthor{\bsnm{Chen}, \binits{Q.}},
\bauthor{\bsnm{Tian}, \binits{L.}},
\bauthor{\bsnm{Waller}, \binits{L.}},
\bauthor{\bsnm{Asundi}, \binits{A.}}:
\batitle{Transport of intensity phase retrieval and computational imaging for partially coherent fields: The phase space perspective}.
\bjtitle{Opt. Lasers Eng.}
\bvolume{71},
\bfpage{20}--\blpage{32}
(\byear{2015})
\end{barticle}
\endbibitem

\bibitem[\protect\citeauthoryear{Khan et~al.}{2021}]{Khan2021-vz}
\begin{barticle}
\bauthor{\bsnm{Khan}, \binits{R.}},
\bauthor{\bsnm{Gul}, \binits{B.}},
\bauthor{\bsnm{Khan}, \binits{S.}},
\bauthor{\bsnm{Nisar}, \binits{H.}},
\bauthor{\bsnm{Ahmad}, \binits{I.}}:
\batitle{Refractive index of biological tissues: Review, measurement techniques, and applications}.
\bjtitle{Photodiagnosis Photodyn. Ther.}
\bvolume{33}(\bissue{102192}),
\bfpage{102192}
(\byear{2021})
\end{barticle}
\endbibitem

\bibitem[\protect\citeauthoryear{Yu et~al.}{2021}]{yu2021plenoctrees}
\begin{bchapter}
\bauthor{\bsnm{Yu}, \binits{A.}},
\bauthor{\bsnm{Li}, \binits{R.}},
\bauthor{\bsnm{Tancik}, \binits{M.}},
\bauthor{\bsnm{Li}, \binits{H.}},
\bauthor{\bsnm{Ng}, \binits{R.}},
\bauthor{\bsnm{Kanazawa}, \binits{A.}}:
\bctitle{Plenoctrees for real-time rendering of neural radiance fields}.
In: \bbtitle{Proceedings of the IEEE/CVF International Conference on Computer Vision},
pp. \bfpage{5752}--\blpage{5761}
(\byear{2021})
\end{bchapter}
\endbibitem

\bibitem[\protect\citeauthoryear{Li et~al.}{2011}]{mdck_paper}
\begin{barticle}
\bauthor{\bsnm{Li}, \binits{J.}},
\bauthor{\bsnm{Gonzalez}, \binits{J.}},
\bauthor{\bsnm{Walker}, \binits{D.}},
\bauthor{\bsnm{Hersom}, \binits{M.}},
\bauthor{\bsnm{Ealy}, \binits{A.}},
\bauthor{\bsnm{Johnson}, \binits{S.}}:
\batitle{Evidence of heterogeneity within bovine satellite cells isolated from young and adult animals}.
\bjtitle{Journal of animal science}
\bvolume{89},
\bfpage{1751}--\blpage{7}
(\byear{2011})
\doiurl{10.2527/jas.2010-3568}
\end{barticle}
\endbibitem

\bibitem[\protect\citeauthoryear{Brassard et~al.}{2021}]{brassard2021recapitulating}
\begin{barticle}
\bauthor{\bsnm{Brassard}, \binits{J.A.}},
\bauthor{\bsnm{Nikolaev}, \binits{M.}},
\bauthor{\bsnm{H{\"u}bscher}, \binits{T.}},
\bauthor{\bsnm{Hofer}, \binits{M.}},
\bauthor{\bsnm{Lutolf}, \binits{M.P.}}:
\batitle{Recapitulating macro-scale tissue self-organization through organoid bioprinting}.
\bjtitle{Nature Materials}
\bvolume{20}(\bissue{1}),
\bfpage{22}--\blpage{29}
(\byear{2021})
\end{barticle}
\endbibitem

\end{thebibliography}

\end{document}